\documentclass[aps,reprint,nofootinbib,nobibnotes,notitlepage,superscriptaddress,prd,
 amsmath,amssymb
]{revtex4-2}

\usepackage[caption=false]{subfig}
\usepackage{braket}
\usepackage{float}
\usepackage{lipsum}
\usepackage{graphicx}
\usepackage{dcolumn}
\usepackage{bm}
\usepackage{natbib}
\usepackage{hyperref}
\usepackage[capitalise]{cleveref}
\hypersetup{
	colorlinks = true,
    linkcolor = Red,
    urlcolor  = Red,
    citecolor = Red
}

\usepackage{microtype}

\usepackage{verbatim}
\usepackage[amssymb]{SIunits}
\usepackage{tabularx}
\usepackage[dvipsnames]{xcolor}
\usepackage{wasysym}

\usepackage{tikz}

\def\be{\begin{equation}}
\def\ee{\end{equation}}

\usepackage{color}
\definecolor{darkgreen}{RGB}{0,120,0}

\definecolor{darkgreen}{RGB}{0,120,0}

\newcommand{\hMpc}{h\,\mathrm{Mpc}^{-1}}

\newcommand{\av}[1]{\left\langle{#1}\right\rangle} 

\newcommand{\vk}{\vec k}

\newcommand{\vp}{\vec p}

\newcommand{\vq}{\vec q}

\def\beq{\begin{eqnarray}}
\def\eeq{\end{eqnarray}}

\let\vec\mathbf

\makeatletter
\newlength{\apb@width}
\newcommand{\autoparbox}[2][c]{\settowidth{\apb@width}{#2}\parbox[#1]{\apb@width}{#2}}

\makeatother

\def\mnras{\rm{MNRAS}}

\def\aap{\rm{A\&A}}

\def\apj{\rm{ApJ}}                 
\def\apjl{\rm{ApJ}}                
\def\apjs{\rm{ApJS}}               
\def\jcap{\rm{JCAP}}

\begin{document}

\title{The Impact of Non-Gaussian Primordial Tails on Cosmological Observables}

\author{William R. Coulton}
\email{wrc27@cam.ac.uk}
\affiliation{Kavli Institute for Cosmology Cambridge, Madingley Road, Cambridge CB3 0HA, UK}
\affiliation{DAMTP, Centre for Mathematical Sciences, University of Cambridge, Wilberforce Road, Cambridge CB3 OWA, UK}

\author{Oliver H.\,E. Philcox}
\email{ohep2@cantab.ac.uk}
\affiliation{Center for Theoretical Physics, Department of Physics, Columbia University, New York, NY 10027, USA}
\affiliation{Simons Foundation, New York, NY 10010, USA}

\author{Francisco Villaescusa-Navarro}
\email{fvillaescusa@flatironinstitute.org}
\affiliation{Center for Computational Astrophysics, 160 5th Avenue, New York, NY 10010, USA}

\begin{abstract} 
\noindent Whilst current observational evidence favors a close-to-Gaussian spectrum of primordial perturbations, there exist many models of the early Universe that predict this distribution to have exponentially enhanced or suppressed tails. In this work, we generate realizations of the primordial potential with non-Gaussian tails via a phenomenological model; these are then evolved numerically to obtain maps of the cosmic microwave background (CMB) and large-scale structure (LSS). In the CMB maps, our added non-Gaussianity manifests as a localized enhancement of hot and cold spots, which would be expected to contribute to $N$-point functions up to large $N$. Such models are indirectly constrained by \textit{Planck} trispectrum bounds, which 
restrict the changes in the temperature fluctuations to $\mathcal{O}(10\mu\mathrm{K})$.  
In the late-time Universe, we find that tailed cosmologies lead to a halo mass function enhanced at high masses, as expected. Furthermore, significant scale-dependent bias in the halo-halo and halo-matter power spectrum is also sourced, which arises from the squeezed limit of large $N$-point functions that are implicitly generated through the enhancement of the tails. These results underscore that a detection of scale-dependent bias alone cannot be used to rule out single field inflation, but can be used together with other statistics to probe a wide range of primordial processes.
\end{abstract}

\maketitle

\section{Introduction}\label{sec:intro}
\noindent Primordial non-Gaussianity (PNG) represents a pivotal probe of the early Universe. Extensive theoretical investigations have shown that the degree and form of deviation from Gaussianity can encode information about the microphysics of the early Universe model \citep[e.g.][]{2003JHEP...05..013M,Creminelli:2004yq,Silverstein:2003hf,Chen:2009zp}, which can be uniquely accessed through studies of the Gaussianity of the early Universe. As such, PNG is one of the most promising means to extract new information about the physics governing inflation and beyond.

Constraints on primordial non-Gaussianity have primarily been wrought from the cosmic microwave background (CMB; \citep[e.g.,][]{2011ApJS..192...18K,2016A&A...594A..17P,2023PhRvL.131r1001P}) and large scale structure (LSS; \citep[e.g.,][]{Cabass:2022ymb,DAmico:2022gki,2022PhRvL.129b1301C,Cabass:2022oap}) datasets. These searches can be divided into blind searches for \textit{any} non-Gaussianity \citep[e.g.,][]{1998ApJ...503L...1F,2000ApJ...533..575B,2014A&A...571A..23P,Philcox:2023xxk} and targeted searches \citep[e.g.,][]{2005ApJ...634...14K,2015arXiv150200635S}, with the latter providing constraints that are an order of magnitude more stringent, though restricted to a subset of the model space. Most targeted searches have constrained perturbative non-Gaussianity -- \textit{i.e.} constraints on the amplitudes of primordial bispectra and trispectra (higher-dimensional analogs of the skewness and kurtosis). Interesting exception are \citep{Munchmeyer_2019,Philcox_2024}, which performed searches for non-Gaussianity from large $N$-point functions in WMAP data, albeit optimized for particular physical signatures. 

Recent theoretical work has highlighted that many models of inflation can source a non-perturbative spectrum of non-Gaussianities. Mechanisms generating this include non-attractor phases of inflation 
\citep[e.g.,][]{2022JCAP...12..034C,2023PhRvL.131a1002P,2023arXiv230905750D,Biagetti:2018pjj}, inflationary preheating \citep[e.g.,][]{Bond:2009xx,Barnaby:2009wr}, multifield inflation \citep[e.g.,][]{2019arXiv190602827P,2022JHEP...05..052A,Panagopoulos:2019ail} including heavy particles with time dependent masses \citep[e.g.,][]{2017JCAP...10..058F}, features in the inflationary potential landscape \citep[e.g.,][]{Chen:2018uul} including light scalar fields and infrared dynamics \citep[e.g.,][]{Gorbenko:2019rza,Mirbabayi:2019qtx}, non-perturbative inflation treatments including quantum diffusion \citep[e.g.,][]{2020JCAP...03..029E,2022JCAP...02..021T}, dissipation \citep[e.g.,][]{Celoria:2021vjw,Creminelli:2023aly}, and eternal inflation from self-interactions \citep[e.g.,][]{Cohen:2021jbo}, as well as slow reheating phenomena \citep[e.g.,][]{Padilla:2024iyr}. A key focus of many of these investigations has been how heavy tails impact the formation of primordial black holes. If such processes (which we hereafter refer to as ``tail non-Gaussianity'') are at play, however, they could also (or instead) give impacts on larger scales, including those observed in the CMB and LSS datasets. This naturally raises the question: how are the CMB and late-time structures impacted by modifications to the tails of the primordial distribution?

\citep{Munchmeyer_2019,Philcox_2024} explored the signatures of inflationary heavy particle production on the CMB (described in \citep{2017JCAP...10..058F}) and showed that the principal effect was to source spherically symmetric point-source-like features in the primordial fluctuation field. Our work builds on this both by exploring other tailed models and by exploring their effects on the LSS, thus providing the first tailed $N$-body simulations. \citep{Ezquiaga_2022} used the Press-Schechter \citep{Press_1974} formalism to perform the first study of how tail non-Gaussianity changes the halo mass function. This paper develops these ideas further: we complement these analytical methods with N-body simulations and extend the analysis to encompass a broad range of cosmological observables. 

In this work, we implement a phenomenological model that exponentially enhances the tails of the primordial potential probability density function (hereafter PDF). This model can reproduce features of commonly considered non-perturbative PNG models, for example, exponential tails. Notably, we assume a symmetric model, which identically modifies positive and negative tails, and thus cannot generate $N$-point functions with odd $N$. Furthermore, the model is constructed such that the bulk of the PDF remains Gaussian, which is required to reproduce CMB and LSS observations.

\vskip 4pt
The remainder of this paper is structured as follows: in \cref{sec:ICs} we consider the construction of non-Gaussian primordial perturbations and discuss the key properties of our model. In \cref{sec:simulations} we describe how to generate realizations of the CMB and LSS from non-Gaussian initial conditions, before presenting the impacts of tail non-Gaussianity on CMB and LSS observables in \cref{sec:CMB}\,\&\,\cref{sec:LSS}. We present our conclusions in \cref{sec:conclusions}.

\section{Initial Conditions}\label{sec:ICs}

\noindent Theories of the early Universe provide predictions for the statistics of the primordial potential, $\Phi(\mathbf{x})$. Whilst there are many models that predict exponentially enhanced tails, the specific form of the primordial potential PDF depends strongly on the detailed microphysics. That said, some features are fairly generic; for example, many models exhibit tails that scale as $\exp[-C\Phi^N]$ for some $N>0$. Thusly, we here consider the following phenomenological model for the real-space primordial PDF: 
\begin{align}\label{eq:PDF_ICs}
    \Phi(\mathbf{x}) \sim\begin{cases}
         A \exp\{-C\Phi^N\} \,\, \text{if } |\Phi|>\alpha \sigma, \\
         B \exp\{-\frac{\Phi^2}{2\sigma^2}\} \,\,\text{if }|\Phi|\leq\alpha \sigma.
    \end{cases}
\end{align}
Near the center of the distribution $\Phi$ follows a Gaussian PDF with a variance $\sigma^2$, which enforces consistency with observations (that usually do not probe the distribution tails). The tails are modified such that, beyond $\alpha \sigma$, they exhibit a new power-law scaling with exponent $N$. The constants $A$, $B$ and $C$ normalize the distribution, with $A$ and $C$ set by assuming the PDF and its derivative to be continuous at the switch point. For $N<2$ the tails of the distribution are enhanced relative to a Gaussian, whilst for $N>2$ they are suppressed. As discussed in \cref{sec:GenerationMethod}, we enforce that different locations, $\mathbf{x}$ and $\mathbf{x'}$, are spatially correlated so that the power spectrum of the primordial potential matches observations. 

We consider three different models in this work:
\begin{itemize}
    \item $N=1.0$ and $\alpha =2.5$. This model has tails $\sim \exp\{-C\Phi\}$, which are a feature of many models \citep[e.g.,][]{2020JCAP...05..022A,2020JCAP...03..029E,2021PhLB..82036602B,2023PhRvL.131a1002P,2023arXiv230905750D}.
    \item $N=1.5$ and $\alpha =2$. This type of tail non-Gaussianity is seen in \citep{2021JCAP...06..051C}.
    \item $N=2.5$ and $\alpha =2$. Unlike the previous models, this model has suppressed tails compared with a Gaussian. Whilst suppressed tails are less common than thicker tails, they can be produced by a range of mechanisms \citep[e.g.][]{2022JCAP...12..034C,2023JCAP...09..023H} and represent an interesting phenomenological case to explore.
\end{itemize}
We refer to each model by the value of $N$. The value of $\alpha$ was chosen to generate approximately similar levels of non-Gaussianity in each model, as measured via the distribution's kurtosis. As discussed in \cref{sec:CMB}, the above PDFs are already ruled out via CMB trispectrum constraints; however, we adopt these levels to allow clear exploration of observable signatures without the need for enormous and computationally expensive suites of simulations. 

Finally, our model for the primordial tails is symmetric, thus equally impacting positive and negative tails, with a vanishing skewness (as discussed above). Whilst many models predict asymmetric tails, choosing a symmetric distribution allows us to separate effects caused by the tails from effects induced by the skewness (for example from the local $f_{\rm NL}$ parametrization). As such, the leading order impact of the tails will be on the primordial trispectrum, with contributions also expected in all higher $N$-point functions with even $N$.

\subsection{Generation Methodology}\label{sec:GenerationMethod}

\begin{figure}
  \centering
  \includegraphics[width=.5\textwidth]{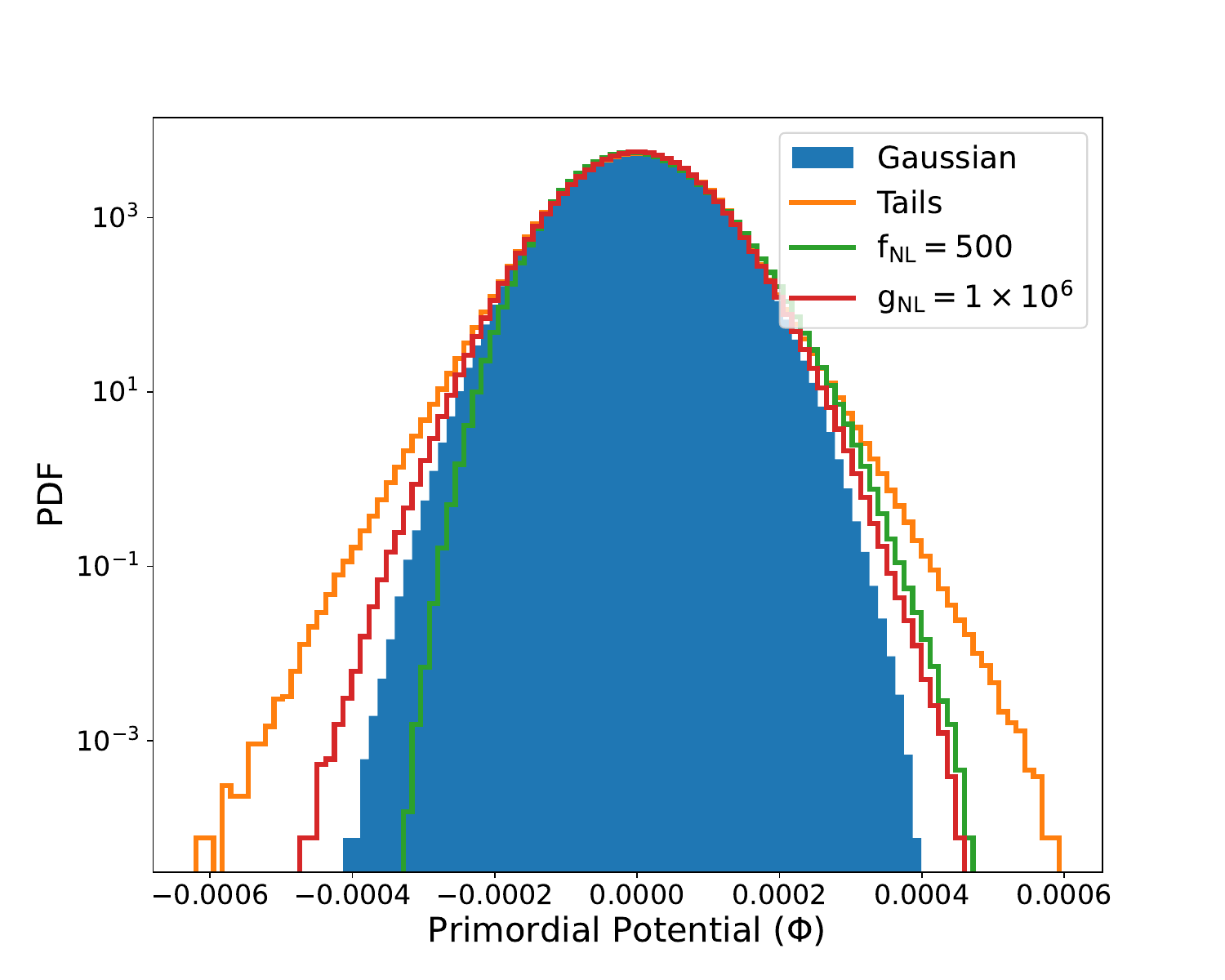} 
\caption{Histograms of a realization of the primordial potential for Gaussian initial conditions (blue), canonical local $f_\mathrm{NL}$ and $g_{\rm NL}$ non-Gaussianity (green and red), and one of the enhanced tail models considered in this work (orange), here with $N=1$. This illustrates the qualitative difference between the different types of non-Gaussianity; even though our model closely matches that of $g_{\rm NL}$ at small $\Phi$, we find significant deviations in the tails.
}
\label{fig:pdf_hist}
\end{figure}

\noindent We require our initial conditions to have two properties: (1) the one-point PDF should be given by \cref{eq:PDF_ICs}; (2) the simulations should have a power spectrum that matches observations, \textit{i.e.}\,$P_\Phi(k)\propto k^{n_s-4}$ where $n_s$ is the spectral index. To achieve this, we use a two-step procedure. First, we generate a real-space Gaussian random field using an input power spectrum, $P_\alpha(k)$, where the specific functional form is detailed below. Second, we convert the PDF to the form in \cref{eq:PDF_ICs} via inverse transform sampling. Specifically, for $\Phi(\mathbf{x})$ at pixel $\mathbf{x}$, we use the cumulative distribution function (CDF) of the Gaussian distribution to compute the probability of having a $\Phi$ value smaller than or equal to the observed value. Next, we use the inverse CDF of the tails distribution, computed from \cref{eq:PDF_ICs}, to map this probability onto a $\Phi$ value under our new PDF. This operation is performed point-wise across the simulation box. The second operation will enforce the correct PDF, however, it will distort the power spectrum from $P_\alpha(k)$. In our approach, we choose $P_\alpha(k)$ such that after the inverse sampling transform, the power spectrum is given by $P_\Phi\propto k^{n_s-4}$, recovering property (2) above. As the level of non-Gaussianity we use is small, the distortions to the power spectrum induced by the inverse sampling transform are also small, thus we can compute the desired $P_\alpha(k)$ in an iterative manner. For this, we start from a guessed power spectrum of $P_0 \propto k^{n_s-4}$, then generate a Gaussian realization, perform the inverse sampling transform, and measure the power spectrum. The ratio of this power spectrum to the input gives a transfer function, $t_0(k)$, which sets the power spectrum of the next iteration; $P_1(k) = P_0(k)/t_0(k)$. The procedure is repeated $n$ times, stopping when the resulting transfer function, $t_n$, is consistent with unity.

\begin{figure}
  \centering
  \includegraphics[width=.5\textwidth]{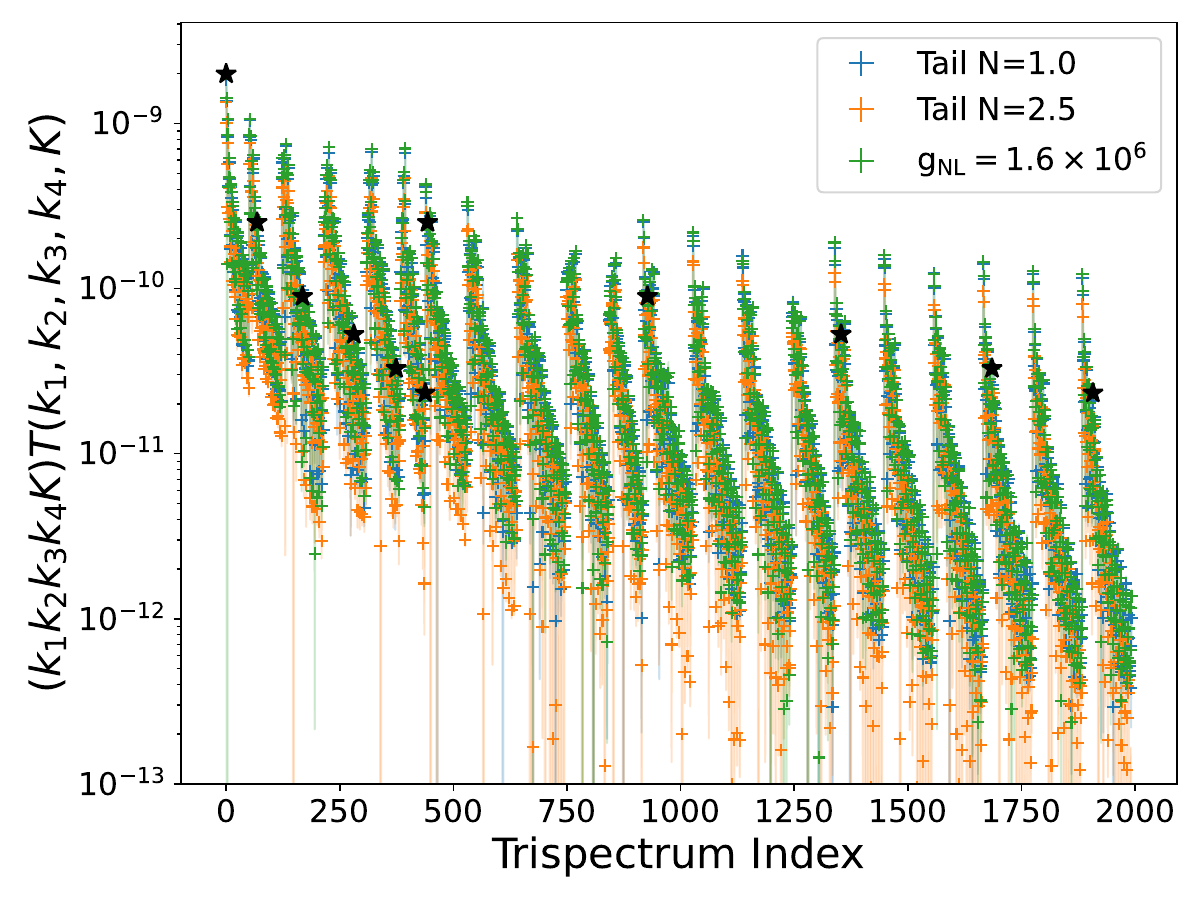} 
\caption{A comparison of the trispectrum of the primordial potential for a model with $g_\mathrm{NL}= 1.64 
\times 10^6$ and two of the enhanced tail models considered here. The squeezed trispectrum configuration, denoted by the stars, are highly similar between the tails and $g_\mathrm{NL}$ models; however, larger differences are seen in the other configurations. The transparent regions denote the $1-\sigma$ error bars.}
\label{fig:primordial_trispectrum}
\end{figure}

\subsection{Theoretical Expectations}

\noindent In \cref{eq:PDF_ICs}, we construct a non-Gaussian distribution for the primordial potential, $\Phi$, by performing a spatially-local transformation $\Phi_G(\mathbf{x})\to \Phi[\Phi_G(\mathbf{x})]$, where $\Phi_G$ is a Gaussian random field. Assuming the effect of tails to be relatively small, this can be represented by its Taylor expansion, \textit{i.e.}\ our formalism is equivalent to rewriting
\beq\label{eq: Phi-transformation}
    \Phi(\mathbf{x}) = \Phi_G(\mathbf{x})\left[1+\sum_{n=1}^\infty X_{\rm NL}^{(n+1)}\Phi_G^{n}(\mathbf{x})\right]
\eeq
(ignoring effects arising from the renormalization of the power spectrum). Here, $X_{\rm NL}^{(2)}$ and $X_{\rm NL}^{(3)}$ are the usual local $f_{\rm NL}$ and $g_{\rm NL}$ amplitudes\footnote{Throughout this paper we use $f_\mathrm{NL}$ and $g_\mathrm{NL}$ to refer to the amplitude of the local bispectrum and trispectrum, respectively. }, and, since our PDF is symmetric, all $X_{\rm NL}^{(n+1)}$ with odd $n$ vanish. As such, our tail model can be thought of as a (possibly non-perturbative) extension to local non-Gaussianity, which sources all higher cumulants. 

The transformation described above sources a number of non-trivial polyspectra: for example, $X_{\rm NL}^{(n)}\Phi^n_G(\mathbf{x})$ gives rise to the following connected $(n+1)$-point function at leading order
\beq\label{eq: n+1-pt}
    \av{\Phi(\vk_1)\cdots\Phi(\vk_{n+1})}'_c &=& n!\,X_{\rm NL}^{(n)}\left[P_\Phi(k_1)\cdots P_\Phi(k_n)\right]\nonumber\\
    &&\,+\,n\text{ perms.}
\eeq
dropping the Dirac delta for brevity. This implies that our model will source a primordial trispectrum, pentaspectrum, quintaspectrum, nonaspectrum, and beyond. Assuming approximate scale-invariance, these spectra have non-trivial squeezed limits:
\beq
    &&\lim_{q\to 0}\frac{\av{\Phi(\vk_1)\cdots\Phi(\vk_n)\Phi(\vq)}'_c}{P_\phi(q)}\\\nonumber
    &&\qquad= n!\,X_{\rm NL}^{(n)}P_\phi(k_2)\cdots P_\phi(k_{n})+(n-1)\text{ perms.},
\eeq
which is a manifestation of the broken cosmological consistency relations. 

\subsection{Properties of the simulated initial conditions}\label{sec:ic_properties}
\noindent For a simple validation of our procedure, we can examine the PDF of the primordial potential, which is shown in \cref{fig:pdf_hist}. For comparison, we plot the PDF obtained from Gaussian initial conditions, as well as those corresponding to quadratic and cubic local-type non-Gaussianity, $f_{\rm NL}$ and $g_{\rm NL}$ (which source a particular primordial bispectrum / skewness and trispectrum / kurtosis). It can be clearly seen that our approach primarily alters the tails of this distribution, somewhat akin to $g_{\rm NL}$, but with a different asymptotic behavior. Note that, by construction, the generation procedure generates a standard primordial power spectrum $P_\Phi(k)\propto k^{n_s-4}$. 

In \cref{fig:primordial_trispectrum} we show measurements of the primordial trispectrum. This is computed using modes between $k=0.01\hMpc$, and $k=0.15\hMpc$ with six equally spaced bins per axis and is estimated from 80 sets of simulated initial conditions, using the same estimators and analysis tools as \citep{Coulton_2023c}. Matching our theoretical expectations, we find that the induced primordial trispectrum exhibits a strong squeezed limit similar to $g_\mathrm{NL}$-type local PNG, which is similar for all choices of tail index $N$. This arises in inflationary models with cubic non-linearities, such as $\Phi(\mathbf{x})=\Phi_\mathrm{G}(\mathbf{x})+g_\mathrm{NL}\Phi^3_\mathrm{G}(\mathbf{x})$, and produces a primordial trispectrum that has a large squeezed configuration (when one of the $k$ modes is small and the others are large), and a PDF shown in \cref{fig:pdf_hist}. Notably, the shape of the trispectrum outside the squeezed limit is quite different. That the tails model produces a squeezed trispectrum (and beyond) is of particular observational consequence, since this will lead to scale-dependent bias in late-time observables. We return to this later in the paper. 
\section{Simulations}\label{sec:simulations}
\noindent Once we have generated non-Gaussian initial conditions, we use established methods to create realizations of the CMB \citep{Liguori_2003,Elsner_2009} and LSS \citep{Scoccimarro:2011pz,Wagner_2012,Coulton_2023}. In this section, we briefly review these methods.

\subsection{CMB simulations}\label{sec:cmb_sims}
\noindent Our method to generate CMB maps is a small modification to the method developed in \citep{Liguori_2003} and \citep{Elsner_2009}. This method has three steps: first we generate Gaussian realizations of the primordial potential in a spherical-radial basis, $\Phi_{\ell m}(r)$. Then we transform the Gaussian realizations into realizations with tail non-Gaussianity, via the method described in \cref{sec:GenerationMethod}. Finally, the radial shells are combined with appropriate weights to obtain a CMB map. These steps are described in detail below.

The spherical harmonic coefficients of the CMB anisotropies, $a_{\ell m}$, can be related to the primordial potential by
\begin{align}\label{eq:a_to_phi}
    a^X_{\ell m} = 4\pi (-i)^\ell \int\frac{\mathrm{d}^3\mathbf{k}}{(2\pi)^3}\Phi(\mathbf{k}) \Delta^X_\ell(k)Y^{*}_{\ell m}(\hat{\mathbf{k}}),
\end{align}
where $\Delta^X_\ell(k)$ are the CMB transfer functions for $X \in \{T,E\}$ (assuming vanishing $B$-modes, given the scalar initial conditions). The challenge with simulating CMB realizations is the large range of scales that contribute to the integral in \cref{eq:a_to_phi}. To avoid simulating very dense and memory-intensive grids, \citep{Liguori_2003} suggested the following approach. First, one expresses the potential in the spherical harmonic basis as
\begin{align}
    \Phi_{\ell m}(k) = \int \mathrm{d}^2\hat{\mathbf{k}}\,\Phi(\mathbf{k})Y_{\ell m}(\hat{\mathbf{k}})
\end{align}
and then utilize the Rayleigh expansion to transform these to real space
\begin{align}
    \Phi_{\ell m}(r) = \int_0^\infty \mathrm{d}k\,k^2 \Phi_{\ell m}(k)j_{\ell}(kr),
\end{align}
where $j_{\ell}(x)$ is the spherical Bessel function of order $\ell$. Generating Gaussian realizations of $\Phi_{\ell m}(r)$ is as straightforward as generating realizations of $\Phi(\mathbf{k})$, since the primordial power spectrum is isotropic and homogeneous. The number of grid points is now set by the number of radial points ($r$) and resolution of the output maps, which sets the $\ell,m$ ranges. Note that it is necessary to simulate the $\ell=1$ modes to correctly reproduce the PDF with the non-Gaussian transform, described below.

Given a Gaussian realization of $\Phi_{\ell m}(r)$, the real-space field $\Phi(\mathbf{x}$) can be obtained via a spherical harmonic transform.  We can then apply the transform, described in \cref{sec:GenerationMethod}, to generate the non-Gaussian field. To obtain a CMB map from this field, we can either invert the steps above or, more simply, use the following relationship
\begin{align}
    a^X_{\ell m} = \int_0^{\infty} \mathrm{d}r\,r^2 \Phi_{\ell m}(r)\Delta^X_\ell(r),
\end{align} where 
\begin{align}
    \Delta^X_\ell(r)= \frac{2}{\pi}\int_0^{\infty} \mathrm{d}k\,k^2 \Delta^X_\ell(k)j_\ell(kr).
\end{align}

As detailed in \citep{Liguori_2003} and \citep{Elsner_2009}, there is a range of optimizations to this algorithm that we make use of to efficiently generate high-resolution ($N_{\rm side}=2048$) \texttt{healpix} maps \citep{gorski_2005}. 

\subsection{Large scale structure simulations}\label{sec:lss_sims}
\noindent Our procedure for generating $N$-body simulations follows closely that of the \textsc{quijote} simulations \citep{Villaescusa-Navarro_2020,Coulton_2023,Coulton_2023c}. First, we evolve the initial conditions to $z=0$ using the linear transfer functions from CAMB \citep{Lewis_2000}, then rescaling to $z=127$ using the linear growth factors (without radiation). The fields at $z=127$ are used to compute the 2LPT displacements from which the initial particle positions are defined. We use the TreePM code \textsc{gadget-3} \citep{Gadget} to evolve these positions and velocities down to $z=0$, saving snapshots at $z=1.0$ and $z=0.0$. Halos are identified with both the Friends-of-Friends (FoF) \citep{FoF} and \textsc{rockstar} \citep{Behroozi_2013} algorithms. The seeds of the initial conditions are chosen to also match the fiducial-cosmology \textsc{quijote} simulations (with simulation indices of 0$-$100), and we adopt the same (standard-resolution) \textsc{gadget} settings. Both of these choices enable seamless comparison between talled simulations and the other cosmologies within the \textsc{quijote} suite. Many estimators of summary statistics required gridded fields; to compute these, we use the Cloud-in-Cell \citep{Hockney_1981} grid assignment scheme, deconvolving the window function from our power spectrum and bispectrum measurements \citep{Jing_2005}.

We generate 100 simulations for each of the three tailed models: $N=1$, $N=1.5$, and $N=2.5$. As seen in \cref{fig:primordial_trispectrum}, these models produce an non-trivial primordial trispectrum. To elucidate which effects arise from the primordial trispectrum and which effects are unique to the tails, we also simulate 100 simulations with $g_\mathrm{NL}=1.64\times 10^6$. This is chosen since (a) $g_\mathrm{NL}$ is a commonly-adopted model of primordial non-Gaussianity that produces squeezed features similar to those of our tails models; (b) the value of $g_\mathrm{NL}$ approximately matches the squeezed limit of the tails models (whose importance is discussed in \cref{sec:halo_field}).
\begin{figure}
    \centering
  \begin{subfloat}[Gaussian, noiseless CMB realization\label{fig:gaus_cmb_sim}]{\includegraphics[width=.45\textwidth]{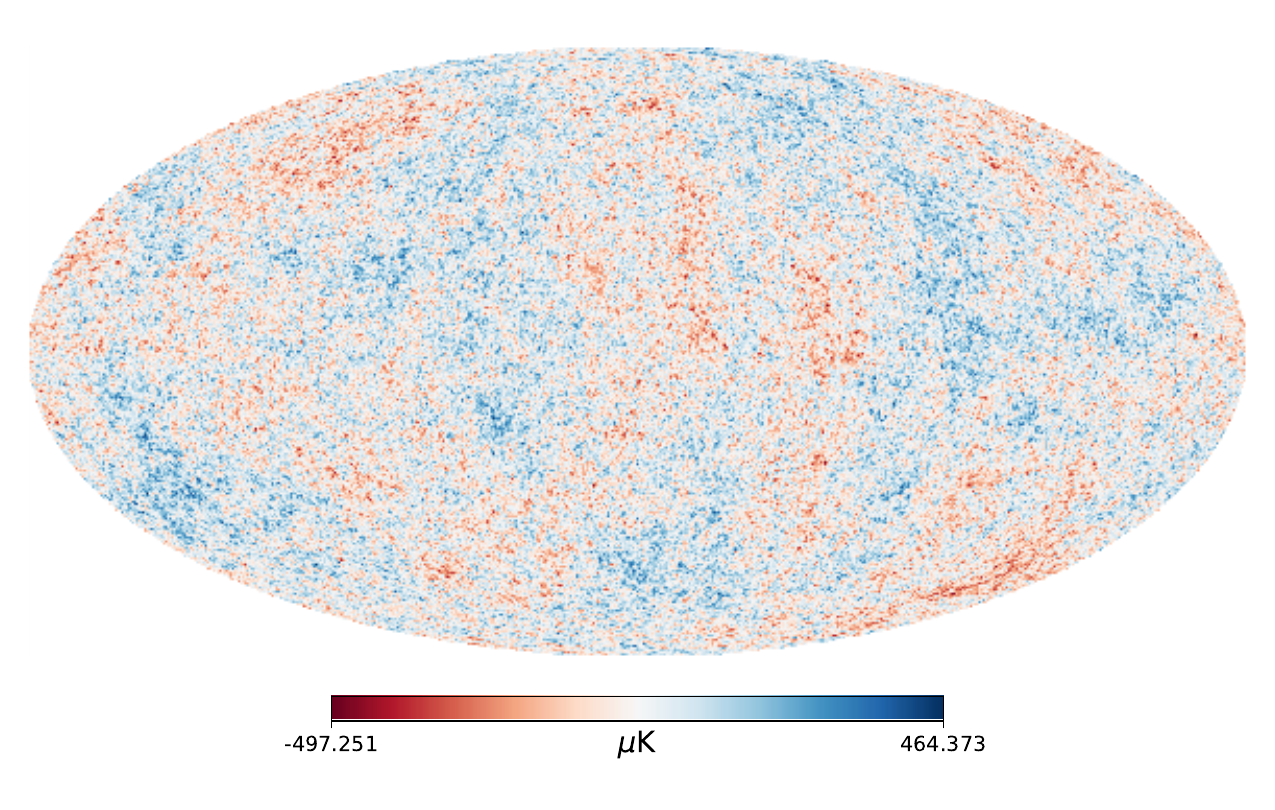}}
  \end{subfloat}
  \begin{subfloat}[$\Delta T$ from tail non-Gaussianity with $N=1$ and $\alpha=2.5$\label{fig:tails_sim_cmb_2p5_1}]
  {\includegraphics[width=.45\textwidth]{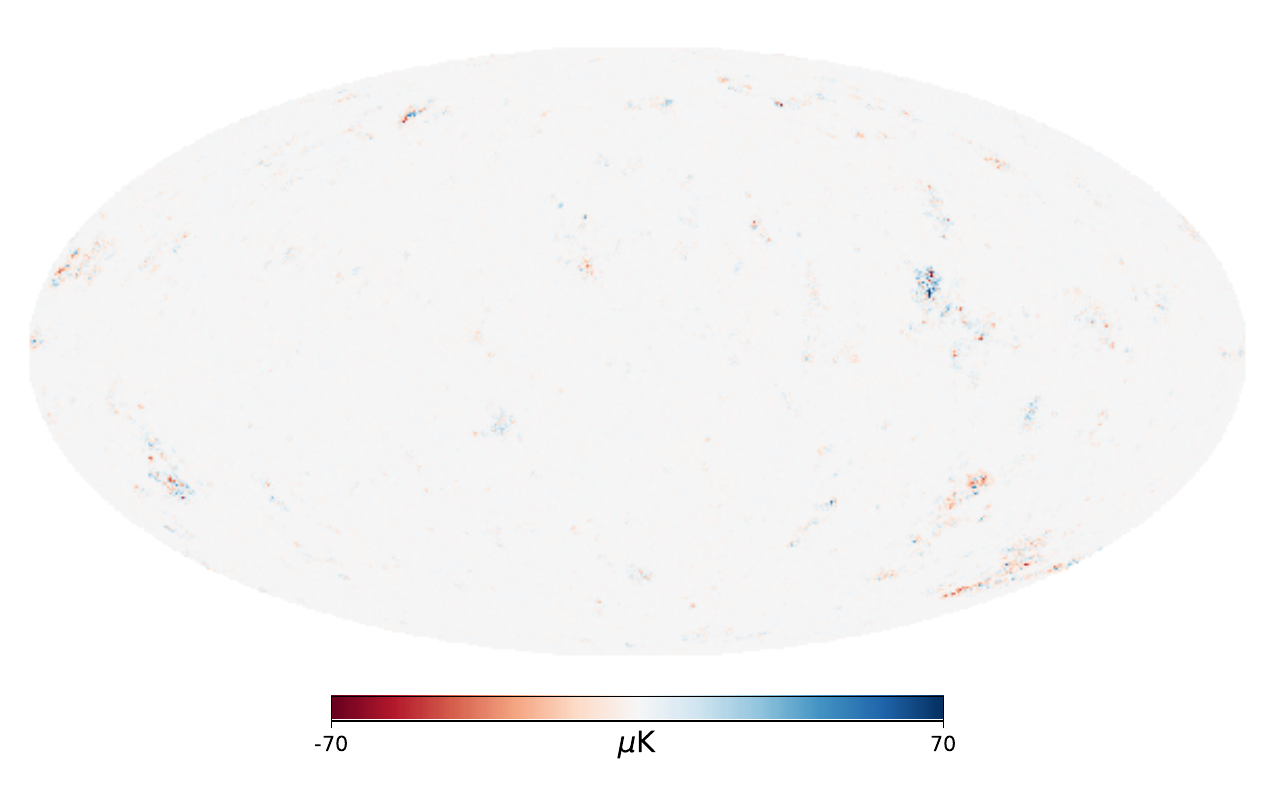}}
  \end{subfloat}
  \\
  \begin{subfloat}[
  $\Delta T$ from tail non-Gaussianity with $N=1$ and $\alpha=3.5$\label{fig:tails_sim_cmb_3p5_1}
  ]{\includegraphics[width=.45\textwidth]{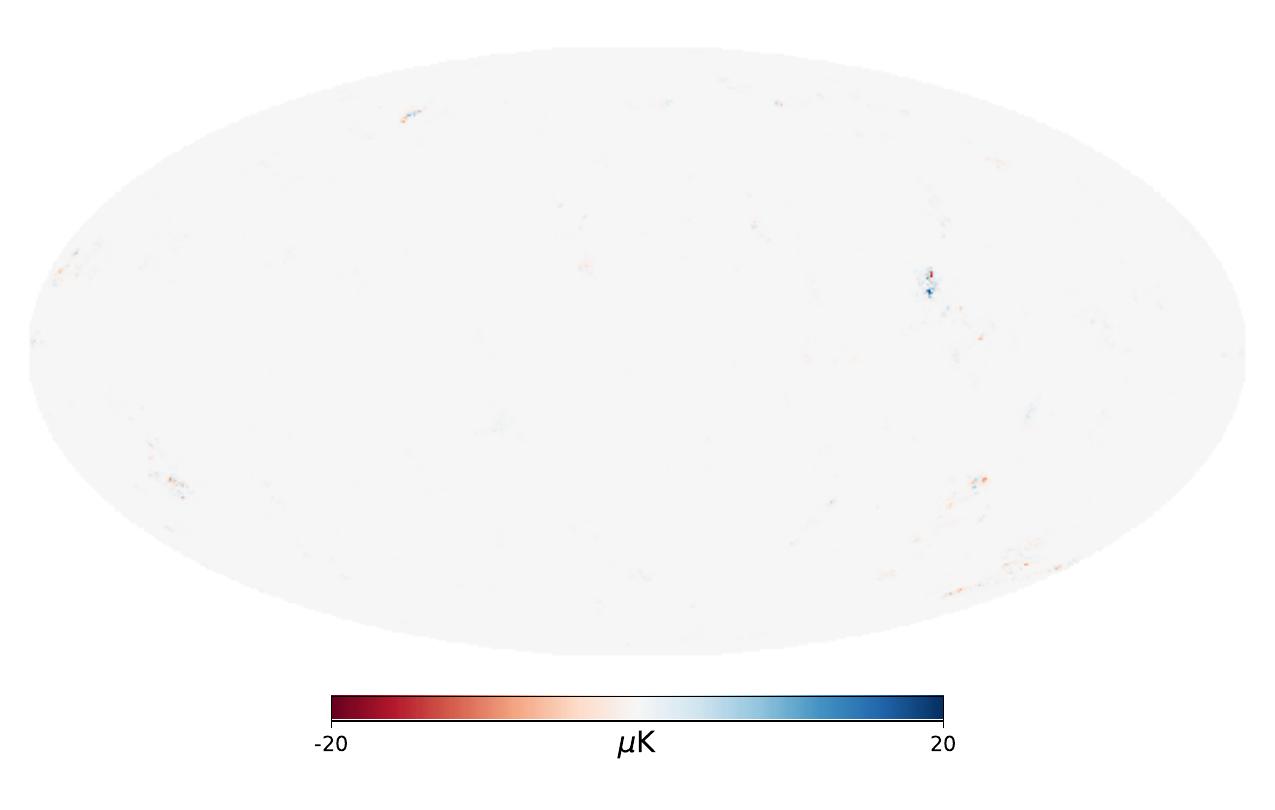}}
  \end{subfloat}
  \caption{Modifications to the CMB temperature maps induced by primordial tails, as parametrized by \cref{eq:PDF_ICs} with two values of $\alpha$. Here, the tail scales as $p(\Phi)\sim e^{-C\Phi}$, for $|\Phi|>\alpha\sigma$, following a Gaussian distribution in the central region. We clearly observe localized structures in the CMB map, which will not be captured by low-order correlation functions. The first set of parameters, $N=1$ and $\alpha=2.5$, correspond to the models analysed in detail here. As discussed in the text, this model is not compatible with CMB trispectrum constraints. The second set, $N=1$ and $\alpha=3.5$, show the impacts on the CMB of a weaker tails model that is not ruled out by past CMB non-Gaussianity measurements.}
    \label{fig:CMB-sims}
\end{figure}

\begin{figure}
    \centering
    \includegraphics[width=.49\textwidth]{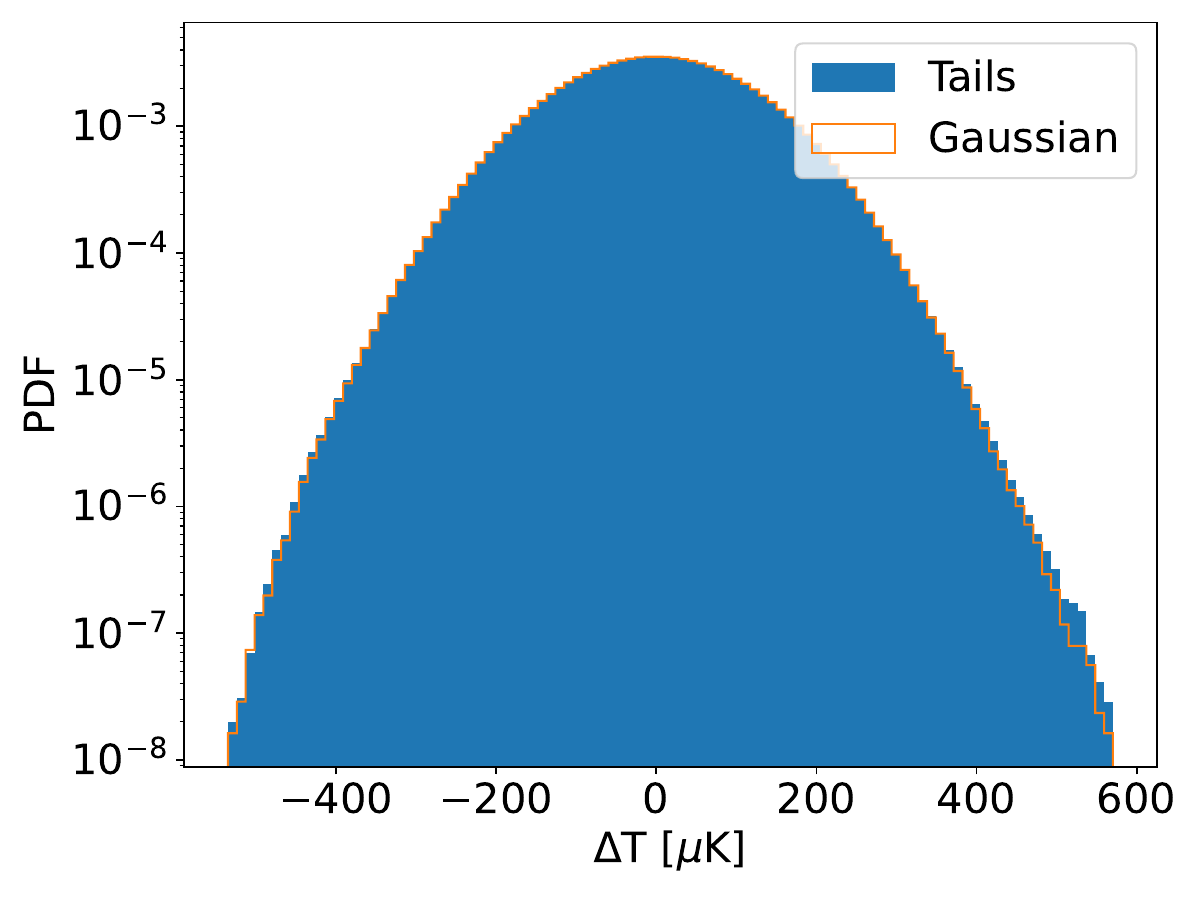}
    \caption{Impact of enhanced tail non-Gaussianity on the CMB PDF, here shown with parameters $N=1$ and $\alpha = 2.5$. Whilst large changes are induced in the 3D primordial PDF, cf.\,\cref{fig:pdf_hist}, this deviation is mostly washed out in the CMB PDF, due to projection effects. A similar conclusion holds for $f_{\rm NL}$ and $g_{\rm NL}$ non-Gaussianity: observationally allowed values give no visible impact on the CMB PDF.}
    \label{fig:cmb-histogram}
\end{figure}

\section{Impact on the cosmic microwave background}\label{sec:CMB}
\noindent First, we assess the impact of tail non-Gaussianity on the CMB. The goal of this investigation is both to understand the phenomenology of tails, and to understand whether they are already highly constrained by observations of CMB non-Gaussianity. For this purpose, we focus on a single model: $N=1$ and $\alpha=2.5$. 

In \cref{fig:CMB-sims}, we show how the noiseless CMB responds to injected tails in the primordial PDF. As our phenomenological model does not change the bulk of the distribution (only rare events with $|\Phi|>\alpha\sigma$), many parts of the CMB map are unaltered; however, some localized features emerge, primarily near hot and cold spots. If one increases the value of $\alpha$ such effects become rarer, as expected. 

Given the large changes to the primordial PDF with tail non-Gaussianity, one may expect to see large changes in the CMB PDF. As seen in \cref{fig:cmb-histogram}, this is not the case. This arises as the CMB is an integrated quantity and combines modes across the surface of last scattering, cf.\,\cref{eq:a_to_phi}, which washes out the signals visible in the 3D potential PDF. 

Despite the limited visibility in the one-point function, the tailed cosmology in question can be well-constrained from higher-order non-Gaussianity. As shown in \cref{sec:ic_properties}, the enhanced tail non-Gaussianity produces a trispectrum with a squeezed limit similar to the cubic local trispectrum, parametrized by $g_{\rm NL}$. The \textit{Planck} satellite constrained {$g_\mathrm{NL}=-5.8\pm 6.5 \times 10^4$ \citep{2020A&A...641A...9P}, whilst for the parameters considered here, we obtain a non-Gaussianity parameter $\hat{g}_\mathrm{NL}=1.1\times10^6$ (see the discussion below), using the methods described in \citep{Philcox_2023,Philcox:2023psd}, adapted to perform a template analysis. As such, the models considered here are already ruled out at the $\sim 20\sigma$ level. As discussed in \cref{sec:ICs}, we choose these levels such that the impact of enhanced tail non-Gaussianity is visible in LSS with a reasonable number of $N$-body simulations.
This is similar to investigations of other types of non-Gaussianity, e.g., \citep{Coulton_2023} simulated (local) $f_\mathrm{NL}=100$, which can be compared to the leading constraint $f_\mathrm{NL}=-0.9\pm 5.1$. From the trispectrum constraint, a slightly modified model with $N=1.0$ and $\alpha=3.5$ is consistent with the data, however, and yields rare features in the temperature maps, as shown in \cref{fig:tails_sim_cmb_3p5_1}. 

Note that the reference $g_\mathrm{NL}$ simulations described above have a different value from that measured from the tailed simulations ($g_{\rm NL}^{\rm ref}=1.64\times 10^6$ versus $\hat{g}_\mathrm{NL}=1.1\times10^6$). This arises since the measured value is obtained by fitting the full $g_\mathrm{NL}$ trispectrum template, whilst the reference value is set by matching the amplitude of the squeezed limits. The former takes a lower value since, outside of the squeezed limit, the tailed and $g_{\rm NL}$-type trispectra have different shapes.

It is also interesting to compare our model to that considered in \citep{Munchmeyer_2019} (see also \citep{Philcox_2024}). Here, we chose a distribution with symmetric tails, which imply vanishing $N$-point functions with odd $N$. As such, the structure of the non-Gaussianity is very distinct from that of \citep{Munchmeyer_2019}, which generates both even and odd $N$-point functions with correlated amplitudes. Despite these differences, the features in the CMB are heuristically similar. This arises as our modifications to the primordial potential are fully local, leading to changes that are similar to the spherical profiles seen in \citep{Munchmeyer_2019}. However, the lack of odd $N$-point functions means our model is likely poorly constrained by the method of \citep{Munchmeyer_2019}, which sums over both odd and even correlators with a common template, practically performing an optimal matched-filter analysis in pixel-space. 

In the literature, there exist a number of works which claim hints of anomalies in the CMB (see \citep{2016CQGra..33r4001S} for a review). The features seen in our maps, \cref{fig:tails_sim_cmb_2p5_1}, are relatively large-scale features, and thus mechanisms such as the above could potentially source explanations for features like the ``southern cold spot''. The large scales are most affected due to the ``averaging" of modes over the surface of last scattering for small scales \citep{2004PhRvD..70h3005B,2021JCAP...04..050K}. Whilst we leave a quantitative analysis of these ideas to future work, the tails models considered herein would likely not explain these anomalies due to the trispectrum constraints.  

\begin{figure*}
  \centering
  \begin{subfloat}[Matter PDF\label{fig:pdf_noRatio_z0}]{\includegraphics[width=.49\textwidth]{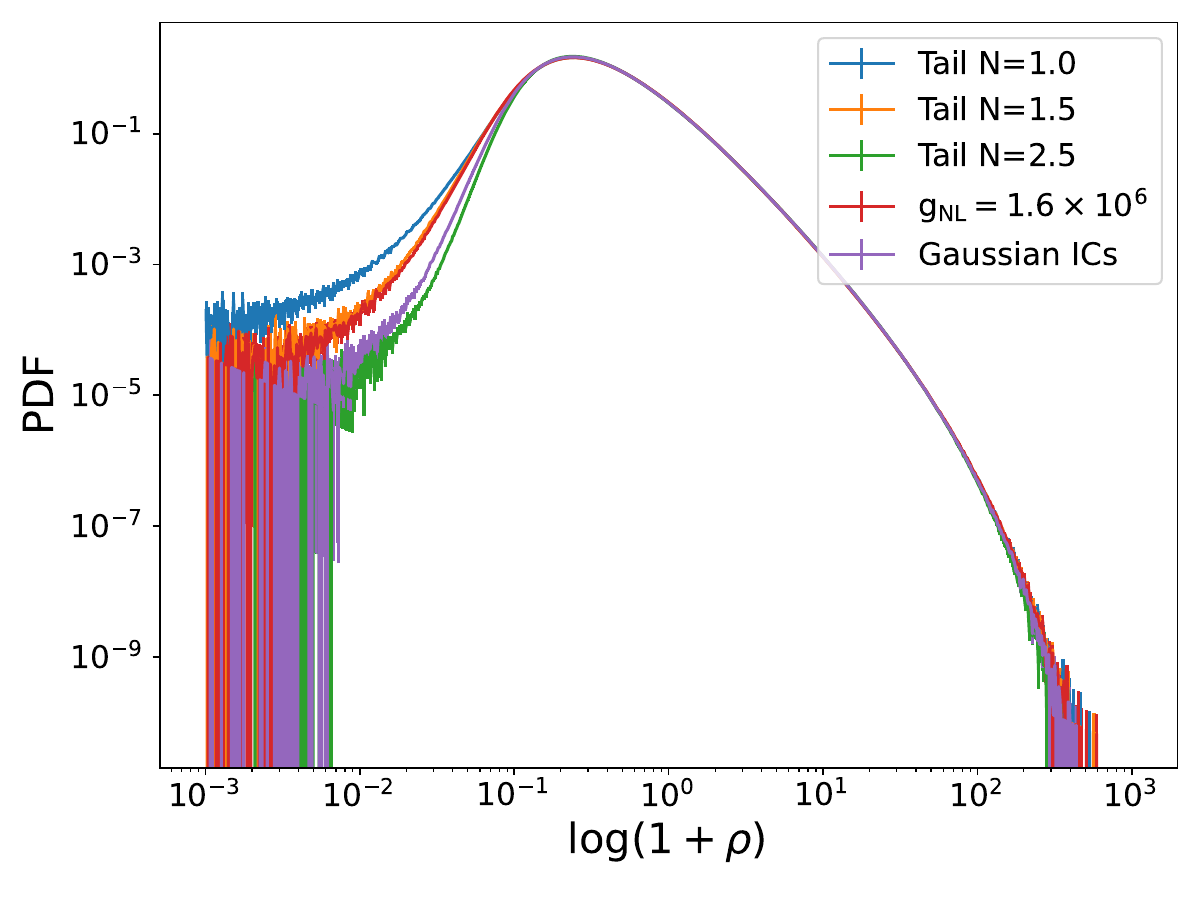} }
  \end{subfloat}
  \hfill
  \begin{subfloat}[Fractional change in the matter PDF\label{fig:pdf_z0}]{\includegraphics[width=.49\textwidth]{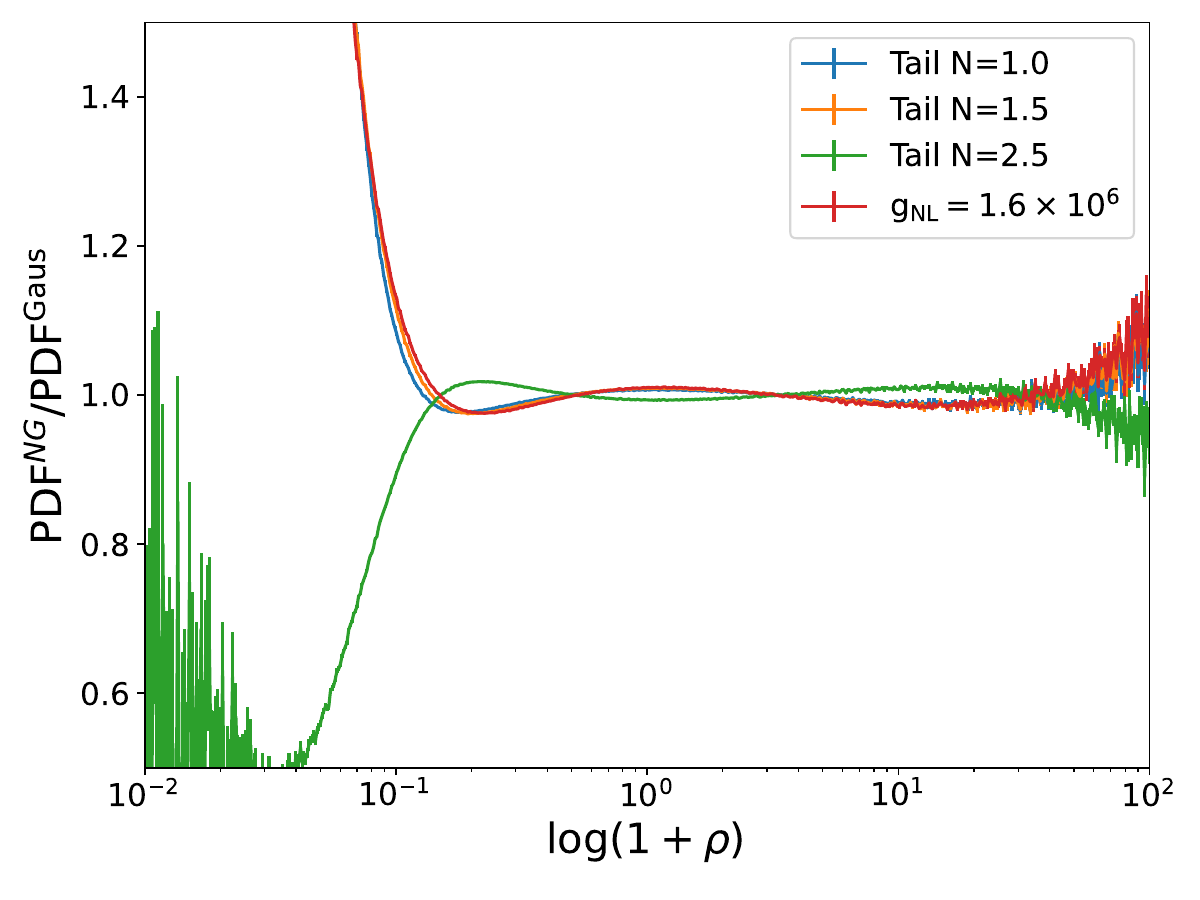} }
  \end{subfloat}
\caption{The matter PDF (left) and the fractional change of the matter PDF from tailed PNG (right), both at redshift zero. For comparison, we show the impact of cubic local non-Gaussianity, with amplitude $g_{\rm NL}=1.64\times 10^6$. The error bars denote the error on the mean of 100 realizations (with significant sample variance cancellation in the right plot). The non-linear evolution of density perturbations generates a highly skewed matter PDF, even for Gaussian initial conditions. However, primordial tail non-Gaussianity still strongly impacts the underdense and overdense regions of the matter field.
}
\label{fig:matter_z0}
\end{figure*}
\section{Impact on large-scale structure statistics}\label{sec:LSS}
\noindent Next, we explore the impact of tails on statistics of the late-time matter (\cref{sec:matter_field}) and halo (\cref{{sec:halo_field}}) density field statistics. 

\begin{figure}
    \centering
\includegraphics[width=.49\textwidth]{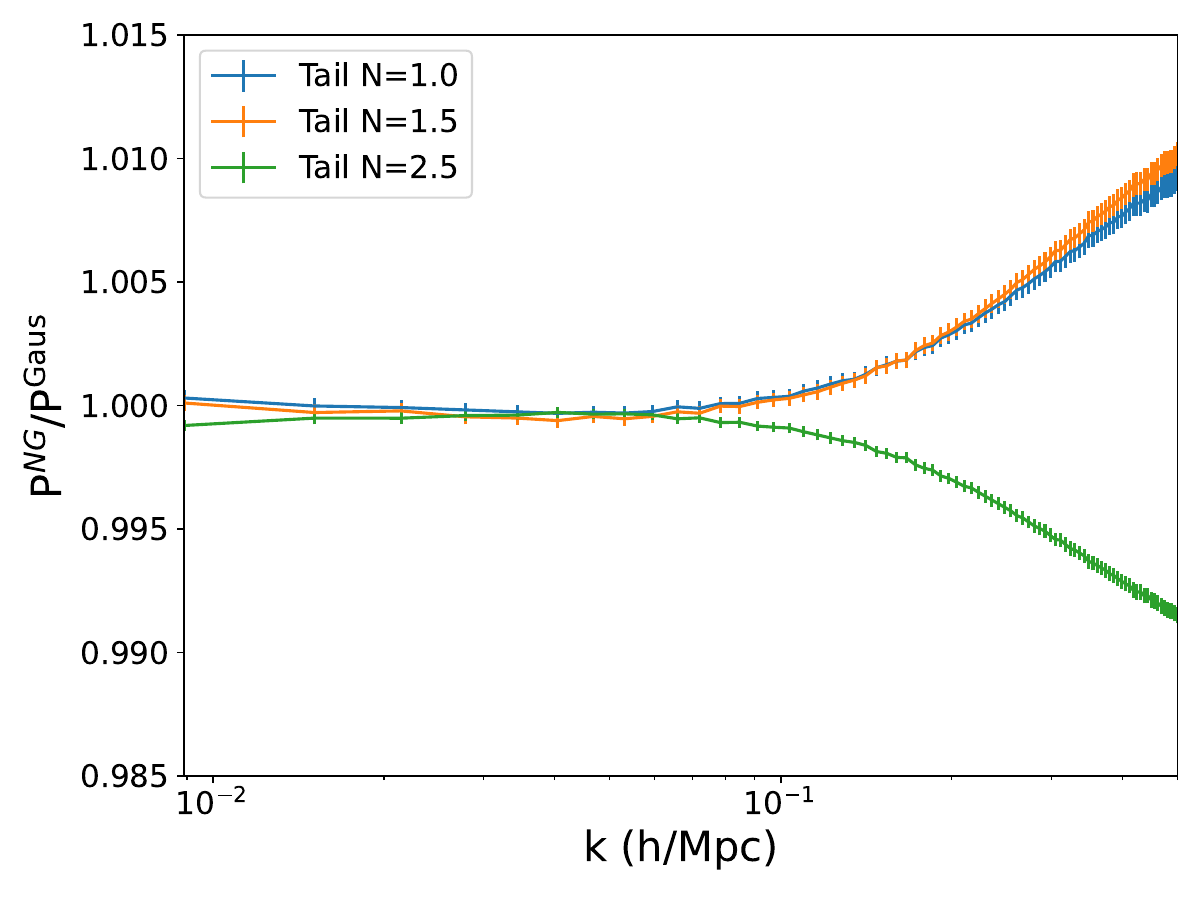}
    \caption{As Fig.\,\ref{fig:matter_z0} but showing the impact on the matter power spectrum at $z=0.0$. The large scale matter power spectrum is essentially unaffected by the addition of primordial tails. This occurs by construction since our primordial tails model has an unchanged primordial power spectrum and thus an unchange linear power spectrum at late-times. On small scales we see that enhanced (suppressed) tails lead to an enhancement (suppression) of the matter power spectrum. The error bars are the error on the mean of the ratio from 100 realizations.
    \label{fig:pk_matter}}
\end{figure}
\begin{figure*}
  \centering
  \begin{subfloat}[Matter Bispectrum \label{fig:bk}]{\includegraphics[width=.49\textwidth]{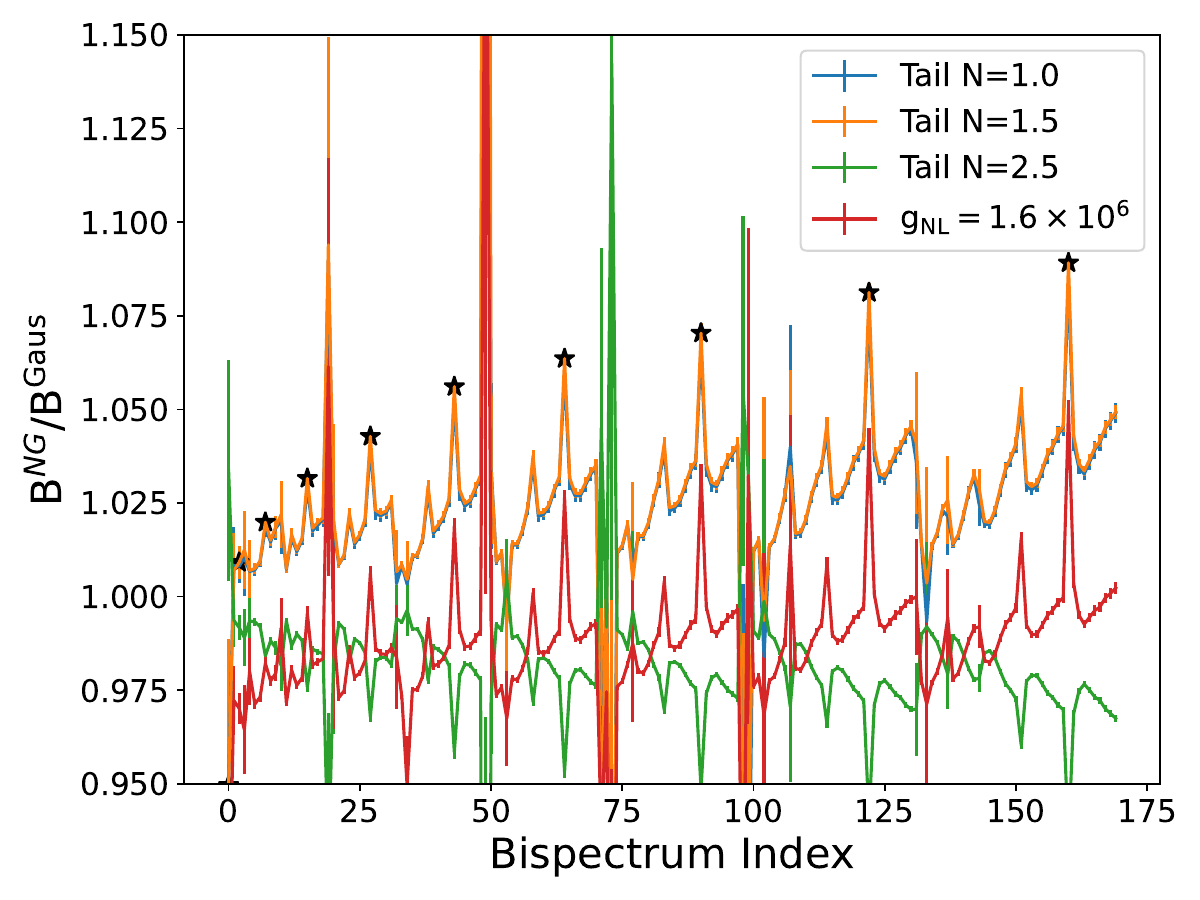} }
  \end{subfloat}
  \hfill
  \begin{subfloat}[Matter Trispectrum \label{fig:tk}]{\includegraphics[width=.49\textwidth]{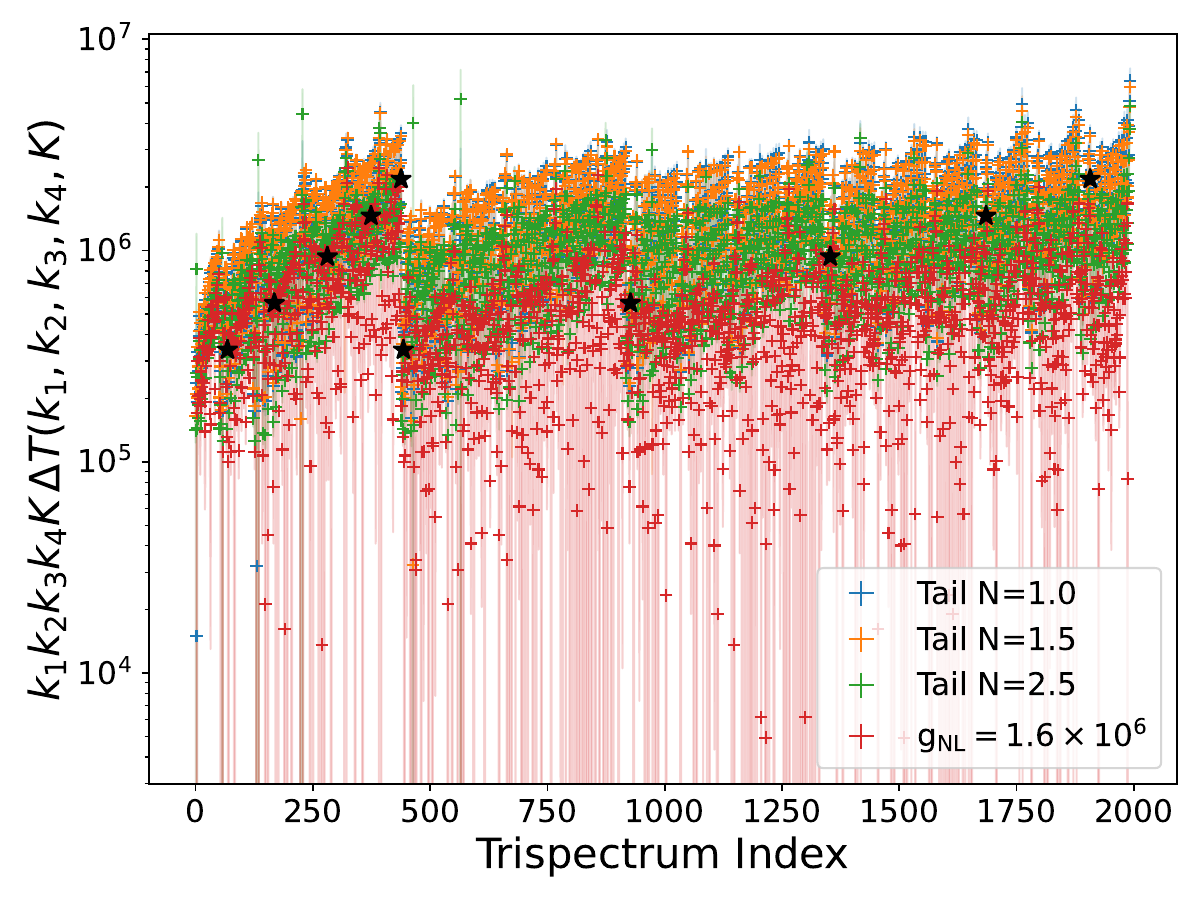} }
  \end{subfloat}
\caption{As Fig.\,\ref{fig:matter_z0} but considering the modification to the matter bispectrum and trispectrum. For the bispectrum, we plot the ratio to the $\Lambda$CDM statistic, whilst we show the difference for the trispectrum, to avoid excess noise. We see find the matter bispectra and trispectra are significantly altered by tail non-Gaussianity. The strongest effects are seen on small scales and in the squeezed limits (when one side corresponds to a large scale mode and the others correspond to small scale modes), which are marked with stars. The error bars on the left hand plot are the error on the mean of the ratio and on the right (semi transparent due the density of points) are the error on the mean of the difference.
}
\label{fig:bk_tk}
\end{figure*}

\subsection{Statistics of the matter field}\label{sec:matter_field}
\subsubsection{Empirical Results}
\noindent All types of tail non-Gaussianity lead to strong impacts on the late-time PDF of matter overdensities, as shown in \cref{fig:matter_z0}. As is clear from \cref{fig:pdf_noRatio_z0}, non-linear evolution from the initial conditions generates a highly asymmetric and non-Gaussian matter PDF even for Gaussian initial conditions. Tail PNG most strongly impacts the extremely underdense and extremely overdense regions (voids and collapsed objects); given that these correspond to the regions of the primordial PDF most impacted by tail non-Gaussianity, this matches expectations. The effect of suppressed tails is almost the mirror opposite, in fractional terms, of the effect of heavier tails, with a suppression at extreme-densities, and an enhancement of densities near the peak of the PDF. In the lowest density regimes, which probe the interior of the voids, the PDF traces the initial conditions and so the suppressed tail model has significantly lower density than the Gaussian case. These results suggest that measurements of the central densities of voids, which are sensitive to the enhancement/suppression of the tails of the primordial PDF, could be a powerful probe of these types of non-Gaussianity.

Changes to the matter power spectrum are shown in \cref{fig:pk_matter}. On large-scales, the matter power spectrum is largely unchanged; this is as expected, since the primordial power spectrum is fixed to that of $\Lambda$CDM, and any contributions from small non-linear scales are parametrically suppressed at low-$k$. The residual low-$k$ differences arise from the imperfect matching of the primordial power spectrum with our iterative procedure (discussed in \cref{sec:GenerationMethod}). On small scales, where the perturbations are no longer linearly related to the primordial anisotropies, we find $\mathcal{O}(1\%)$ modifications. Models that enhance the tails lead to heavier-than-Gaussian tails have enhanced power, whilst lighter tails have suppressed power. 

In \cref{fig:bk} and \cref{fig:tk} we examine the effect of tail non-Gaussianity on the late-time matter bispectrum and trispectrum. The bispectrum is estimated with ten linearly spaced bins between $k=0.01~\hMpc$ and $k=0.20~\hMpc$, whilst for the trispectrum we use six equally spaced bins between $k=0.01~\hMpc$ and $k=0.15~\hMpc$. These are estimated with the standard bispectrum and trispectrum algorithms as implemented in \citep{Foreman_2020,Coulton_2023,Coulton_2023c}. As with the power spectrum, these statistics are enhanced for models with heavier tails and suppressed for models with lighter tails. Generally we find that small scales are impacted strongly. In particular, the squeezed limits of the bispectrum and trispectrum are amongst the most significantly impacted, and we find similar squeezed behavior in both the heavy-tailed cosmologies and the $g_{\rm NL}$ reference simulations.

\subsubsection{Theoretical Expectations}
\noindent Before continuing to discuss halos, we briefly pause to understand the above results theoretically. To begin, note that the matter overdensity field $\delta$ depends non-linearly on the primordial potential $\Phi$ in the late Universe. Writing the linear field as $\delta_L(\vk)\equiv \mathcal{M}_\Phi(k)\Phi(\vk)$ for transfer function $\mathcal{M}_\Phi(k)$ (scaling as $k^2$ on large-scales), the non-linear field can be written as a formal (but not necessarily convergent) perturbative expansion
\beq\label{eq: matter-expansion}
    \delta(\vk) &=& \sum_{n=1}^\infty \int_{\vp_1+\cdots \vp_n=\vec \vk}F_n(\vp_1,\cdots,\vp_n)\\\nonumber
    &&\,\times\,\delta_L(\vp_1)\cdots \delta_L(\vp_n),
\eeq
where $F_n$ encode the equations of motion and are equal to the Standard Perturbation Theory forms (up to renormalization) on sufficiently large scales. For Gaussian initial conditions, the correlators of $\delta(\vk)$ can be obtained from Eq.\,\eqref{eq: matter-expansion} and Wick's theorem; for example, the power spectrum is given by 
\beq\label{eq: matter-Pk}
    P_\delta(k) &=& \sum_{n=1}^\infty\sum_{n+m=\mathrm{odd}}^\infty\int_{\vp_1+\cdots \vp_n=\vk}\int_{\vp_{n+1}+\cdots \vp_{n+m}=-\vk}\\\nonumber
    &&\,\times\,F_n(\vp_1,\cdots,\vp_n)F_m(\vp_{n+1},\cdots,\vp_{n+m})\\\nonumber
    &&\,\times\,\av{\delta_L(\vp_1)\cdots \delta_L(\vp_n)\delta_L(\vp_{n+1})\cdots \delta_L(\vp_{n+m})},
\eeq
where the last line can be written in terms of $(n+m)/2$ power spectra, contracting each combination of pairs of fields. 

In the presence of non-Gaussian initial conditions, the late-time statistics are altered, since the expectation in Eq.\,\eqref{eq: matter-Pk} can also include higher-order contractions, made possible by the non-trivial $N$-point functions shown in Eq.\,\eqref{eq: n+1-pt}. Firstly, late-Universe $N$-point functions can mirror the primordial correlators; for example $X_{\rm NL}^{(3)}\equiv g_{\rm NL}$ sources a trispectrum:
\beq
    \av{\delta(\vk_1)\cdots\delta(\vk_4)}'_c &\supset& 6X_{\rm NL}^{(3)}\frac{\mathcal{M}_\phi(k_4)}{\mathcal{M}_\Phi(k_1)\mathcal{M}_\Phi(k_2)\mathcal{M}_\Phi(k_3)}\\\nonumber
    &&\,\times\,P_L(k_1)P_L(k_2)P_L(k_3)+3\text{ perms.},
\eeq
where the $\mathcal{M}_\Phi$ transfer functions appear due due to the conversion of $\Phi$ to $\delta$ and we write $P_L\equiv \av{\delta_L\delta_L^*}$. In the squeezed limit, this takes the form
\beq
    \lim_{q\to 0}\frac{\av{\delta(\vk_1)\cdots\delta(\vq)}'_c}{P_L(q)} &\supset& 6X_{\rm NL}^{(3)}\frac{\mathcal{M}_\Phi(k_3)}{\mathcal{M}_\Phi(k_1)\mathcal{M}_\Phi(k_2)\mathcal{M}(q)}\\\nonumber
    &&\,\times\,P_L(k_1)P_L(k_2)+2\text{ perms.},
\eeq
with the asymptotic scaling $1/q^2$. This is common to the squeezed limits of all local polyspectra, \textit{i.e.}\ the squeezed $(n+1)$-point function scales as $X_{\rm NL}^{(n)}/q^2$ if $(n+1)$ is even, as observed in the right panel of \cref{fig:bk_tk} for $n=3$.

Secondly, we may source new polyspectra from a coupling of inflationary and late-time non-Gaussianity. For the power spectrum, the lowest-order contribution takes the form
\beq
    P_\delta(k)&\supset& 6X_{\rm NL}^{(3)}\mathcal{M}^2_\Phi(k)\int_{\vp\,\vq}F_2(\vp,\vk-\vp)F_2(\vq,-\vk-\vq)\nonumber\\
    &&\,\times\,\left[P_\phi(p)P_\phi(q)P_\phi(|\vk-\vp|)+\text{3 perms.}\right]
\eeq
and 
\beq
    P_\delta(k)&\supset& 12X_{\rm NL}^{(3)}\mathcal{M}^2_\Phi(k)\int_{\vp\,\vq}F_3(\vp,\vq,\vk-\vp-\vq)\nonumber\\
    &&\,\times\,\left[P_\phi(p)P_\phi(q)P_\phi(k)+\text{3 perms.}\right].
\eeq
Similar expressions can be wrought for higher-order $X_{\rm NL}^{(n)}$ contributions, which require $F_n$ factors with larger $n$. The low-$k$ limits of the SPT kernels imply that, on large scales, the first term scales as $X_{\rm NL}^{(3)}k^4\mathcal{M}_\Phi^2(k)$ whilst the second is asymptotically $X_{\rm NL}^{(3)}\mathcal{M}_\Phi^2(k)k^2P_\phi(k)$, which can be compared to the Gaussian scaling $\mathcal{M}_\Phi^2(k)P_\Phi(k)$. Within $\Lambda$CDM, the second term dominates, thus the fractional corrections to the power spectrum start at $\mathcal{O}(k^2)$. This matches Fig.\,\ref{fig:pk_matter}. A similar conclusion can be wrought at higher-orders: large-scale modifications to the power spectrum are always suppressed by factors of $k^2$.

Due to our symmetric primordial PDF, we expect no contributions to the primordial bispectrum. At late-times, however, this can be broken due to late-time corrections, as above. Considering the lowest-order correction to Gaussianity, $X_{\rm NL}^{(3)}\equiv g_{\rm NL}$, we can form the late-time bispectrum
\beq
    &&B_\delta(\vk_1,\vk_2,\vk_3) \supset 6X_{\rm NL}^{(3)}\mathcal{M}_\Phi(k_1)\mathcal{M}_\Phi(k_2)\mathcal{M}_\Phi(k_3)\nonumber\\
    &&\quad\,\times\,\int_{\vp}F_2(\vp,\vk_1-\vp)P_\Phi(p)P_\Phi(|\vk_1-\vp|)P_\Phi(k_2)\nonumber\\
    &&\quad\,+\,\text{11 perms.},
\eeq
with permutations appearing due to both the choice of quadratic leg (carrying $F_2$) and in the trispectrum of Eq.\,\eqref{eq: n+1-pt}. The squeezed limit of this can be derived straightforwardly:
\beq
    &&\lim_{q\to0}\frac{B_\delta(\vk_1,\vk_2,\vq)}{P_L(q)} \sim X_{\rm NL}^{(3)}\frac{\mathcal{M}_\Phi(k_1)\mathcal{M}_\Phi(k_2)}{\mathcal{M}_\Phi(q)}f(\{k_i\}),
\eeq
for some scalar function $f$; importantly this scales as $1/q^2$ just as for the primordially-sourced poles. The implication of this is that, in the presence of non-linear structure formation and primordial Gaussianities (e.g., from tails), the squeezed limit of the bispectrum will show distinct enhancements, with the same shape as even correlators, but with suppressed amplitudes, since they are generated only by loops (and thus heavily suppressed on large-scales). This matches the conclusion of Fig.\,\ref{fig:bk_tk}.

\subsection{Statistics of dark matter halos}\label{sec:halo_field}

\begin{figure*}
  \centering
  \begin{subfloat}[$z=0.0$\label{fig:hmf_z0}]{\includegraphics[width=.49\textwidth]{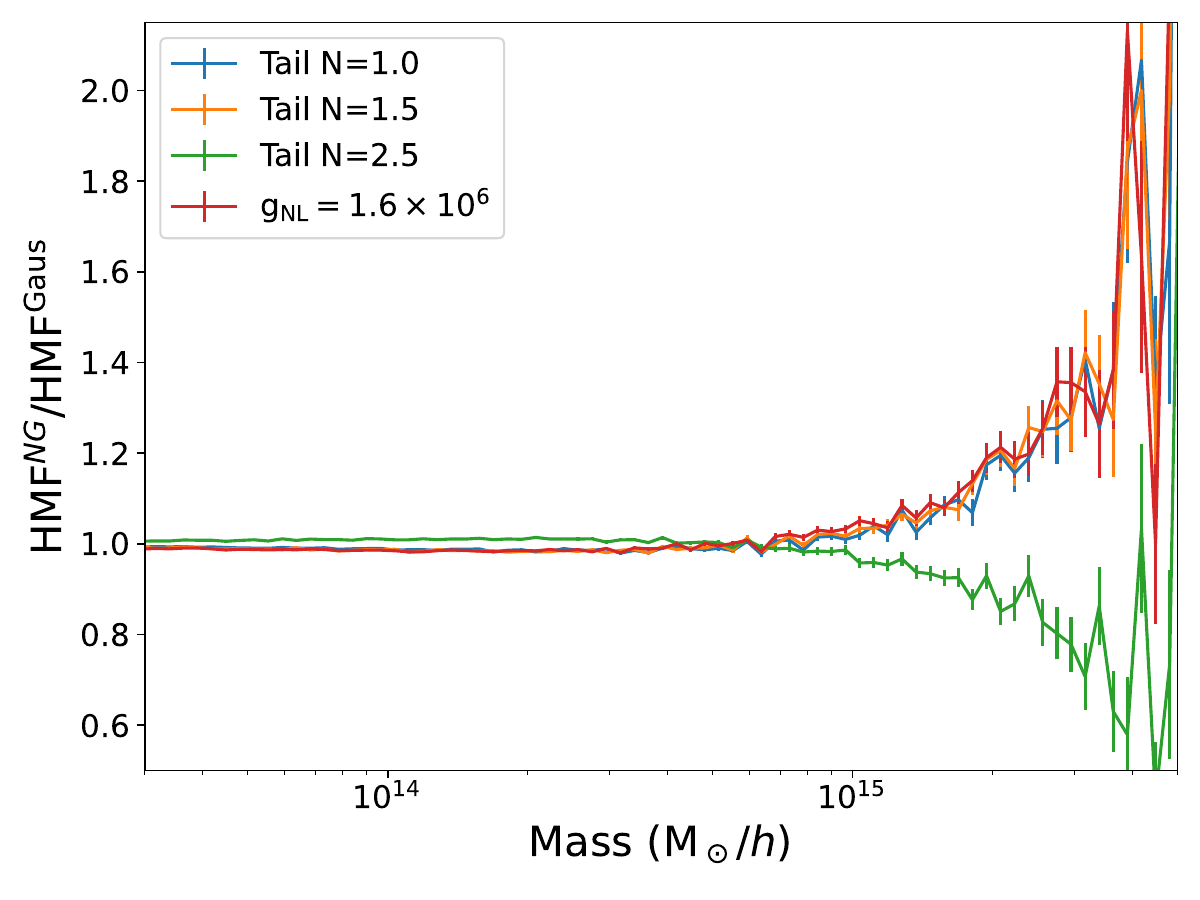} }
  \end{subfloat}
  \hfill
  \begin{subfloat}[$z=1.0$\label{fig:hmf_z1}]{\includegraphics[width=.49\textwidth]{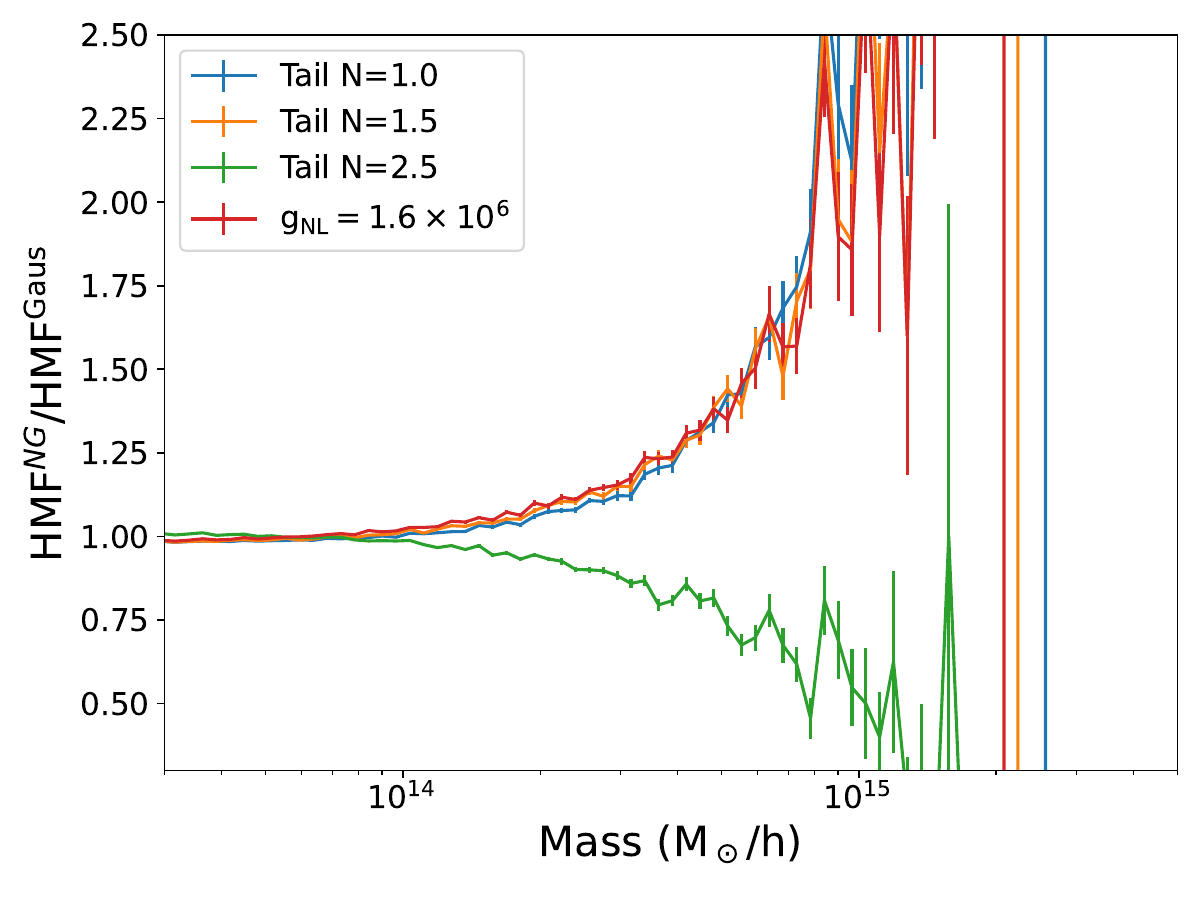} }
  \end{subfloat}
\caption{As Fig.\,\ref{fig:matter_z0}, but showing the fractional change to the halo mass function at $z=0$ (left) and $z=1$ (right). Matching theoretical expectations, an increase (decrease) in the chance of large primordial fluctuations, such as that provided by enhanced (suppressed) tail models, leads to an enhancement (reduction) in the number of massive objects. The size of this enhancement is larger at higher redshift.
\label{fig:hmf} }
\end{figure*}
\begin{figure*}
  \centering
  \begin{subfloat}[$M_\mathrm{halo}>3.2\times 10^{13}h^{-1}M_{\odot}$, $z=0$\label{fig:pk_hh}]{\includegraphics[width=.49\textwidth]{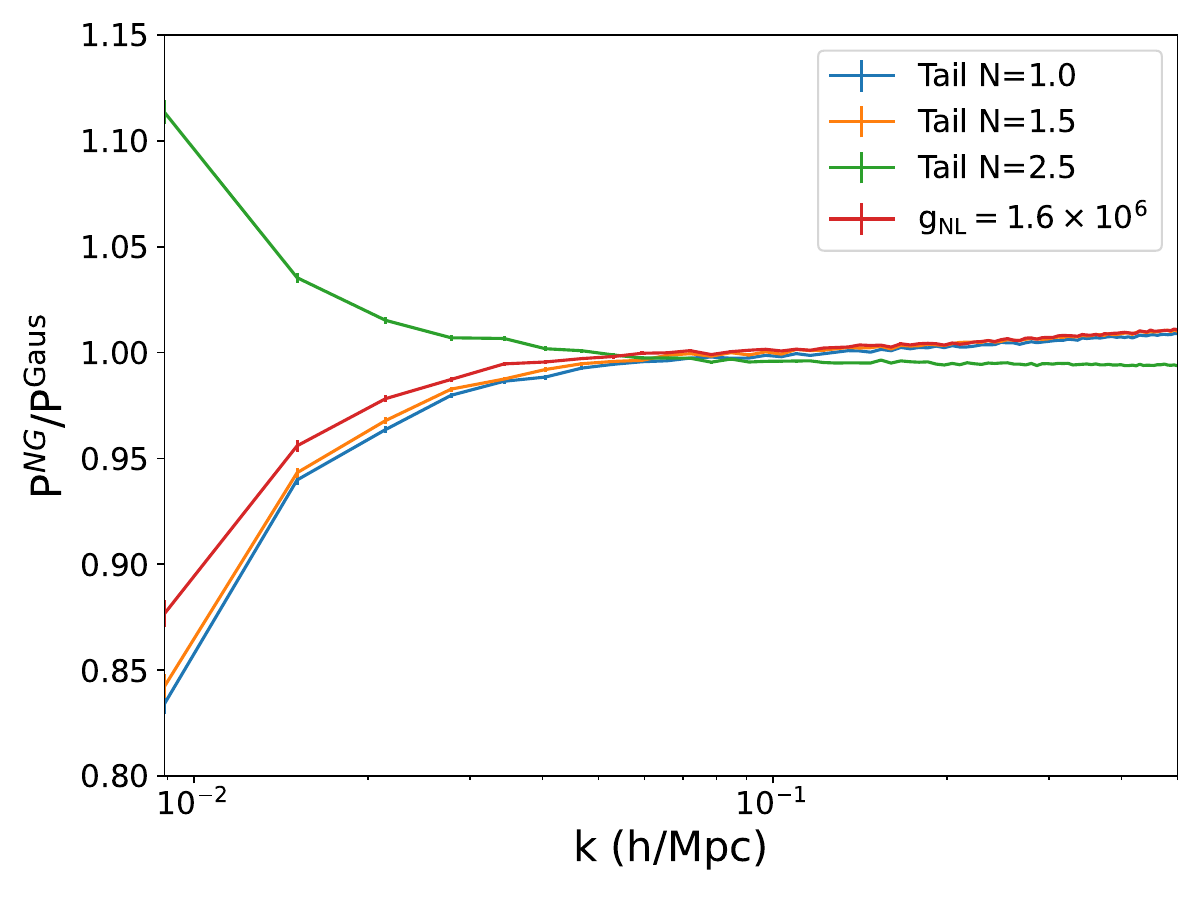} }
  \end{subfloat}
  \hfill
  \begin{subfloat}[Mass and redshift evolution\label{fig:pk_hh_mass_zdep}]{\includegraphics[width=.49\textwidth]{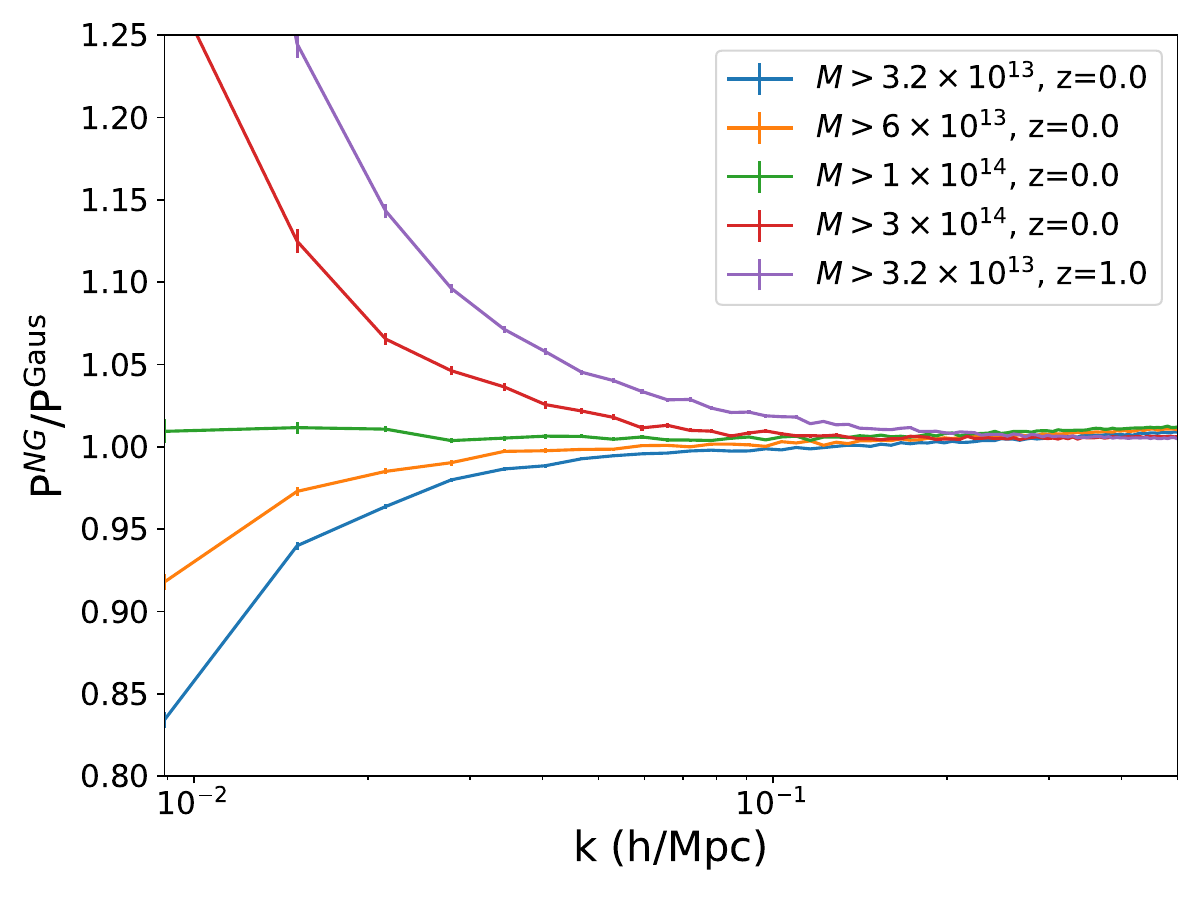} }
  \end{subfloat}
\caption{As Fig.\,\ref{fig:matter_z0} but showing the effect of tails on the halo power spectrum monopole for both a fiducial mass sample (left) and for a range of masses and redshifts (right). All the tails models considered here produce large scale-dependent bias features. This is an interesting result as scale-dependent bias is often considered a ``smoking-gun" of multi-field inflationary models, but tail non-Gaussianity can arise in many single-field models. Notably, the level of scale-dependent bias is larger than for the reference $g_\mathrm{NL}$ case, despite the fact that they have the same (primordial) squeezed-limit trispectra. This implies that higher-order squeezed $N$-point functions contribute to the scale-dependent bias effect. The mass and redshift evolution of this effect would provide one means of differentiating the effect from observational systematics.
}
\label{fig:pk_halo}
\end{figure*}

\begin{figure}
    \centering
    \includegraphics[width=.5\textwidth]{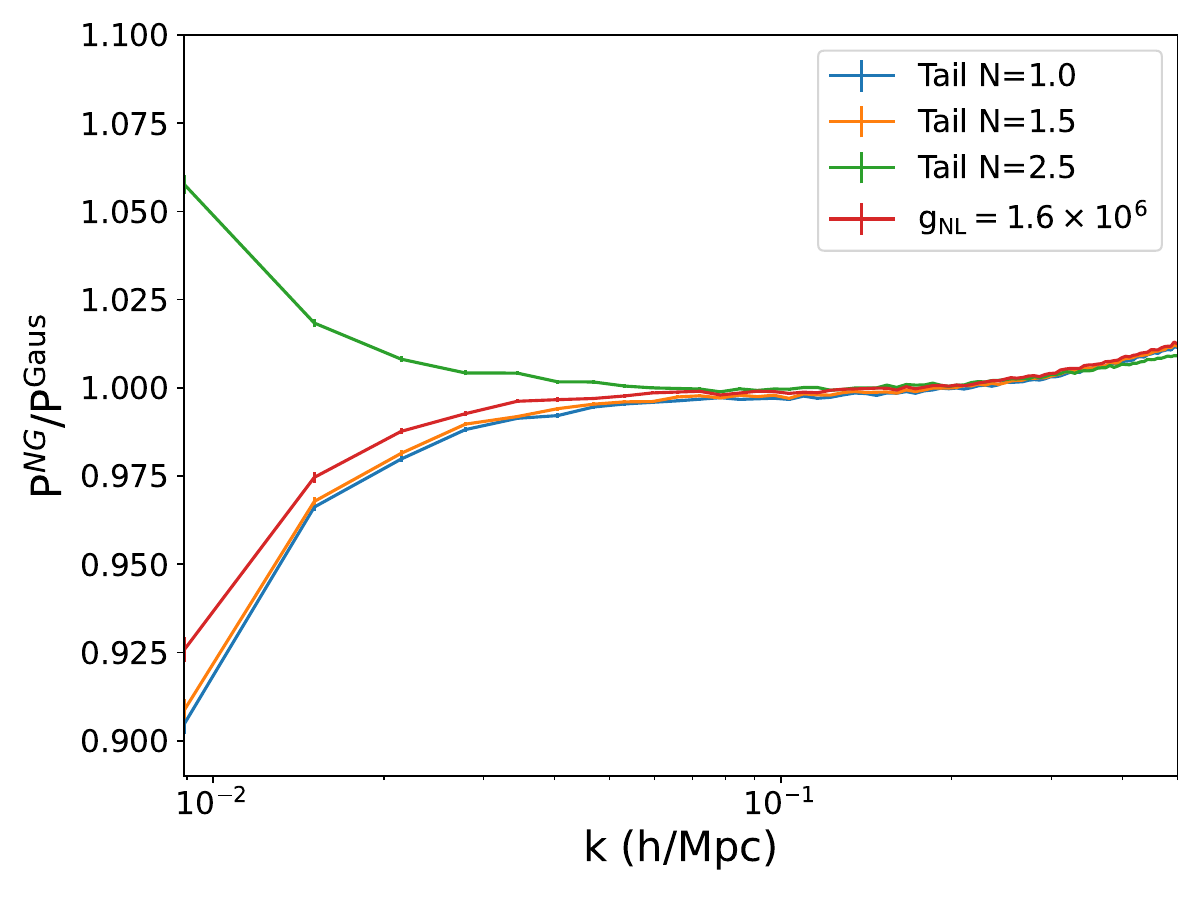}
    \caption{As Fig.\,\ref{fig:matter_z0} but considering the response of the halo-matter cross-spectrum to tail non-Gaussianity. As in the halo auto-power spectrum, we see a clear signature of scale-dependent bias. This allows the effect to be differentiated from primordial processes with collapsed (as opposed to squeezed) $N$-point functions as they would produce scale-dependent bias only in the auto-spectrum. 
    }
    \label{fig:pk_mh}
\end{figure}

\subsubsection{Empirical Results}
\noindent Intuitively, one expects that increasing (decreasing) the tails of the primordial PDF will lead to more (less) massive halos.
This is trivially seen in the Press-Schechter \citep{Press_1974} formalism which relates the number of halos per unit mass ($\mathrm{d}n/\mathrm{d}M$, hereafter the halo mass function or HMF) to the matter probability distribution by
\begin{equation}
\frac{\mathrm{d}n}{\mathrm{d}M} = - \frac{\Omega_m \rho_c}{M\sigma(M)}\frac{\mathrm{d}}{\mathrm{d}M}\left[\int\limits^\infty_{\delta_c}\mathrm{d}\delta P(\delta,M)\right],
\end{equation}
where $\rho_c$ is the critical density, $\sigma(M)$ is the variance smoothed by a top hat sphere that encloses mass, $M$, and $P(\delta,M)$ is the probability distribution for matter perturbations, $\delta$, smoothed by the same sphere. Increasing the tails increases the integrand for high mass objects thus increasing the number of massive objects.

To explore this, we plot the halo mass functions of the tail simulations in \cref{fig:hmf}. In general, the trend is as expected; however, the pattern reverses for low-mass halos. This arises from mass conservation: given the finite amount of matter in the Universe, increasing (decreasing) the number of massive halos must lead to a decrease (increase) in low-mass halos. We find that the changes on the HMF are stronger at higher redshifts, \cref{fig:hmf_z1}, which also matches expectations. Interestingly, whilst at $z=0$ the impact of the two enhanced tail models ($N=1$ and $N=1.5$) are very similar, differences begin to appear at higher redshift, though the effect is small and noisy. To understand this, we note that perturbations in the matter field grow and collapse into halos once they cross a critical threshold. The altered tails mean that there are more (or fewer for suppressed tails) large deviations in the density field, and these cross the collapse threshold earlier than their Gaussian equivalents. See \citep{Ezquiaga_2022} for a derivation of this effect in the Press-Schechter formalism. 

Next, we examine the halo power spectrum monopole. This is shown in \cref{fig:pk_hh}, considering all dark matter halos with $M_h>3.2\times 10^{13}h^{-1} M_\odot$ at redshift zero. Interestingly, on large scales, we see a departure from the Gaussian spectrum that scales as $\sim 1/k^2$. This form is similar to the scale dependence expected from local $f_\mathrm{NL}$ or $g_\mathrm{NL}$ (also plotted) \citep{Dalal_2008,Slosar_2008,Desjacques_2009}, however, the amplitude is larger. The sign and size of the scale-dependent bias change for different halo samples and different redshifts, as seen in \cref{fig:pk_hh_mass_zdep}. Equivalent features are also seen in the halo-matter cross-power spectrum (shown in \cref{fig:pk_mh}), and the halo quadrupole and hexadecapole.

\subsubsection{Theoretical Expectations}
\noindent We now explain the above phenomena mathematically. The transform to generate the modified tails, \cref{eq:PDF_ICs}, is a local operation and, as long as the injected non-Gaussianity is small, can be expressed as a power series of the input Gaussian field at that point, \textit{i.e.}\ terms of the form $\Phi^N(x)$ for integer $N$. Furthermore, such local modifications to the primordial field generate squeezed non-Gaussianity in the $(N+1)$-point function. As derived in \citep{2013JCAP...05..001B,2013MNRAS.435..934F}, large-scale scale-dependent bias is generated from any primordial non-Gaussianity associated with a squeezed $N$-point function, with a leading contribution proportional to $\sim 1/k^2$. Explicitly they show that 
\begin{align}\label{eq:halobiasPrediction1}
    P_\mathrm{halo-halo}(k)&=P_\mathrm{\delta}(k)\left[ b(M)^2+2b(M)\sum\limits_n \alpha_n \kappa_{1,n}(k,M) \right. \nonumber \\ &\left.+\sum\limits_{m,n}\beta_{m,n}\kappa_{n,m}(k,M)\right]
\end{align}
where $b(M)$ is the (Gaussian) halo bias of halos with mass, $M$; $\alpha_{n}$ and $\beta_{n,m}$ are dimensionless coefficients. Finally, $\kappa_{m,n}(k,M)$ is defined as
\begin{align}\label{eq:halobiasPrediction}
    \kappa_{m,n}(k,M)\propto \int d\mathbf{x}\, e^{i\mathbf{k}\cdot(\mathbf{x}-\mathbf{x}')}\langle \delta^m_M(\mathbf{x})\delta^n_M(\mathbf{x}')\rangle,
\end{align}
and encodes squeezed and collapsed $N$-point functions, given a matter density field $\delta_M(\mathbf{x})$ smoothed on the mass scale, $M$.

The outcome of the above is a $1/k^2$ feature in auto- and cross-spectra, which physically arises since the tracer density depends not only on the matter overdensity but also the primordial potential. The latter dependence can be expressed in terms of the matter overdensity and a transfer function that has a $1/k^2$ large scale limit. The same effect arises also in $g_{\rm NL}$-cosmologies; despite the fact that the $g_\mathrm{NL}$ simulations used were chosen to have the same squeezed limit as the tails models (and thus the same trispectrum-induced contribution to scale-dependent bias), we find a large scale-dependent bias effect in the tail models. This can arises due to contributions from higher-order squeezed primordial $N$-point functions, \textit{i.e.} higher values of $n$ in the summation in \cref{eq:halobiasPrediction1}. 

We see very similar features for the two enhanced tail models. The level of non-Gaussianity from the two models was chosen to be approximately similar so it is expected to see qualitatively similar effects, especially as the squeezed trispectra of these models are similar. The small differences seen in these models are almost statistically significant ($\sim3-4\sigma$) and may arise due the difference in higher $N$-point functions. The suppressed tail model shows the opposite behavior, as one would expect. Quantitatively, the effect of the suppressed tails is smaller than the enhanced tails. Given our choice of parameters, the squeezed trispectrum from the suppressed model is slightly smaller than for the enhanced models, which likely explains much of this difference.

An important conclusion of this exercise is that, if a signal of scale-dependent bias were to be detected, it would not be possible to differentiate altered primordial tails from a $f_\mathrm{NL}$ or a $g_\mathrm{NL}$ model using the power spectrum alone or its cross-correlation with (for example) CMB lensing. To differentiate such scenarios, measurements of higher-order $N$-point functions would be needed.

\section{Conclusions}\label{sec:conclusions}
\noindent In this work, we have examined the impact of exponentially enhancing or suppressing the tails of the primordial potential PDF, motivated by the recent interest in models of the early Universe with heavy tails. This form of non-Gaussianity contrasts with most previously studied cases that are characterized only by (scale-dependent) skewness and kurtosis (\textit{i.e.} 3- and 4- point functions). Our study is primarily phenomenological: we have modified the primordial PDF via a heuristic model that captures the enhanced tails whilst leaving the bulk of the PDF unchanged. 

We first investigated the impact of these models on the primary CMB anisotropies, finding that localized large-scale features are generated, as small-scale changes are partially erased by the ``averaging" of anisotropies across the surface of last scattering. Currently, constraints from the \textit{Planck} satellite on the amplitude of the local primordial trispectrum provide the most stringent limit on the tail-induced non-Gaussianities generated in our model \citep{2020A&A...641A...9P}; however, direct searches for heavy tails, possible via model-specific approaches such as \citep{Munchmeyer_2019}, can place more stringent constraints on such models, including non-perturbative information leaking into the tower of higher-point functions. Whilst some of the most significant changes to the CMB maps are large-scale features, the models here are unlikely to provide a consistent explanation for potential large-scale CMB anomalies seen in WMAP and \textit{Planck} data as the size of such effects is practically limited by the CMB $g_{\rm NL}$ bounds.

The non-Gaussianity considered herein strongly impacts the late-time PDF of matter. This could be indirectly observed; extrema counts of weak lensing maps \citep[e.g.,][]{2010PhRvD..81d3519K,2011PhRvD..84d3529Y,2018MNRAS.474..712M,2020MNRAS.495.2531C}, for example, are sensitive to the full matter PDF and thus the presence of tails. The bispectrum and trispectrum of the late-time matter fields are also enhanced (particularly on small-scales and in squeezed limits), with the former effect arising from a coupling of primordial and late-time non-Gaussianity.

As expected, heavier (lighter) tails lead to more (fewer) high-mass halos since the probability of finding a large fluctuation is increased (decreased). As such, altered tails lead to large changes in the halo mass function. Recent works have claimed that James Webb Space Telescope (JWST) observations have found high redshift galaxies with masses larger than expected in current galaxy formation models \citep{2023Natur.616..266L,2023arXiv230810932C,2023NatAs...7..731B}. Whilst this discrepancy likely arises from the uncertainties associated with the formation and evolution of these first galaxies \citep[see e.g.,][]{2023arXiv230404348Y,2023arXiv230413755M}, many works have explored cosmological explanations. For example, \citep{2023ApJ...944..113B} discussed whether local non-Gaussianity could explain these results and 
\citep{2015PhRvD..92b3524C,2015ApJ...814...18H,2023arXiv230611993H,2023MNRAS.526L..63P} explored a blue-tilted primordial power spectrum. Given the astrophysical challenges, we do not explore this dataset in detail in this work; however, we muse that the tail model considered herein would also lead to an enhanced number of massive, high redshift halos, and, if the primordial PDF is symmetric, would evade $f_{\rm NL}$ bounds.

In the presence of tail non-Gaussianity, the halo power spectrum multipoles show a $1/k^2$ feature on large-scales, arising from  scale-dependent bias. This is similar in form to the local $f_\mathrm{NL}$ or $g_\mathrm{NL}$ parametrizations and arises from primordial $N$-point functions with non-trivial squeezed limits (as previously predicted in \citep{2013JCAP...05..001B,2013MNRAS.435..934F}). From this feature alone, it is not possible to differentiate between local type quadratic non-Gaussianity (\textit{i.e.} $f_{\rm NL}$), a primordial trispectrum, or the tails models discussed herein. Unlike collapsed forms of non-Gaussianity such as $\tau_\mathrm{NL}$, local models and our tail simulations also source scale-dependent bias in the matter-halo cross-spectrum. It would be interesting to explore whether recent developments in LSS consistency relations could be used to probe these effects \citep[e.g.][]{Goldstein_2022}.

An important caveat with the above conclusions arises from the following question: do all microphysical tailed models generate squeezed $N$-point functions, or is this simply an artefact from our choice of phenomenological model? Any model whose tails are sourced \textit{locally} would be expected to source such a signature (with non-attractor phases of single field inflation and quantum diffusion providing notable examples \citep[e.g.][]{2013EL....10139001N,2020JCAP...03..029E}), but the conclusions are less clear if the tails are correlated between the potential at far-separated spatial locations.
To completely specify an inflationary model, we would need to not just match the power spectrum and $1$-point function, but also all possible the $N$-point functions (encoding spatial variations); given the infinitude of possible inflationary models, this represents a significant computational challenge.

It is interesting to ask whether we expect CMB or LSS-based observations to provide better constraints on the types of models considered above. Unfortunately, it is difficult to quantify the information content of even the CMB (with LSS being even harder); unlike $f_\mathrm{NL}$ and $g_\mathrm{NL}$, the information is not localized to a single $N$-point function and thus cannot be simply enumerated. To provide a rough estimate of the relative constraining power of the CMB and LSS, we can consider contrasting constraints obtainable from CMB trispectrum measurements with those from scale-dependent halo-bias (a key target for current and future surveys). The tails models considered herein have trispectra that would be detected by \textit{Planck} $g_\mathrm{NL}$ at $\sim 20 \sigma$ (or perhaps more, if one performs a targeted analysis of the tail-induced trispectrum shape, which we find to differ significantly from the standard $g_{\rm NL}$ form). For a sample of dark matter halos with $M>3.2\times 10^{13}~h^{-1}M_\odot$, we find that the scale-dependent bias features would be detectable at $\sim 5 \sigma$ with a $1~h^{-3}\mathrm{Gpc}^3$ survey. Thus a galaxy survey with a volume exceeding $16~h^{-3}\mathrm{Gpc}^3$ would be required to provide tighter constraints than the CMB. This volume is approximately equal to that of the first year of DESI observations, and is thus an rapidly attainable goal \citep{DESI_2024III}. Of course, the exact conclusions depend on the precise galaxy sample and scale-cuts, though the use of sample variance cancellation techniques, either through the use of cross correlations \citep{Schmittfull_2018} or multiple tracers \citep{Seljak_2009,McDonald_2009}, gives further hope for optimism. 

Constraining local-type non-Gaussianity is a key science goal of many current and upcoming experiments \citep[e.g.][]{Schmittfull_2018,dore_2022}, most of which aim to measure it via the scale-dependent bias signature. Given that many single-field models can generate exponentially enhanced tails, this above paper (alongside previous theoretical work) clearly highlights that any detection of scale-dependent bias (or of a non-zero squeezed matter $N$-point function) does not necessarily imply (low-order) local primordial non-Gaussianity, nor does it necessarily rule out single-field inflation. Constraining inflation may thus be a little harder than one expects.

\acknowledgements
We thank Simone Ferraro, Misao Sasaki, Enrico Pajer, Vincent Vennin, Angelo Caravano, Drew Jamieson and Naoki Yoshida for useful discussions. OHEP is a Junior Fellow of the Simons Society of Fellows. The authors are grateful to the organizers of the ``Cosmic Window to Fundamental Physics" conference at the Instituto de F\'{i}sica Te\'{o}rica, Madrid, where this project idea originated. WRC and OHEP authors thank the Bad Moos Saunarium for opening their physical and intellectual pores.
\appendix

\bibliographystyle{apsrev4-1}
\bibliography{refs}

\end{document}